\documentclass[3p]{elsarticle}

\usepackage{color,colortbl}
\usepackage{amsmath}
\usepackage{array}
\usepackage{multirow}
\usepackage{moreverb}
\usepackage{graphicx}
\usepackage{algorithm}
\usepackage{algorithmic}

\RequirePackage{subfigure}[1995/03/06]

\usepackage[dvips,colorlinks,bookmarksopen,bookmarksnumbered,citecolor=red,urlcolor=red]{hyperref}
\usepackage{color}
\usepackage{todonotes}

\usepackage{amssymb}

\biboptions{longnamesfirst,comma}

\journal{Building and Environment}

\begin{document}

\begin{frontmatter}



\title{A Gaussian Process Emulator Approach for Rapid Contaminant Characterization with an Integrated Multizone-CFD Model \tnoteref{t1}}

\tnotetext[t1]{This work was supported in part by Basic Science Research Program through the National Research Foundation of Korea (NRF) funded by the Ministry of Education, Science and Technology (2010-0025484), and in part by the KI Project through KAIST Institute for Design of Complex Systems.
}

\author[kaist]{Piyush M.~Tagade\fnref{fn1}}
\ead{piyush.tagade@kaist.ac.kr}

\author[kaist]{Byeong-Min~Jeong\fnref{fn2}}
\ead{bmjeong@lics.kaist.ac.kr}

\author[kaist]{Han-Lim~Choi\corref{cor1}\fnref{fn3}}
\ead{hanlimc@kaist.ac.kr}

\cortext[cor1]{Corresponding Author}
\fntext[fn1]{Postdoctoral Research Fellow}
\fntext[fn2]{M.S. Student}
\fntext[fn3]{Assistant Professor}

\address[kaist]{Division of Aerospace Engineering, KAIST, Daejeon 305-701, Republic of Korea}

\begin{abstract}
This paper explores a Gaussian process emulator based approach for rapid Bayesian inference of contaminant source location and characteristics in an indoor environment.
In the pre-event detection stage, the proposed approach represents transient contaminant fate and transport as a random function with multivariate Gaussian process prior.
Hyper-parameters of the Gaussian process prior are inferred using a set of contaminant fate and transport simulation runs obtained at predefined source locations and characteristics.
This paper uses an integrated multizone-CFD model to simulate contaminant fate and transport.
Mean of the Gaussian process, conditional on the inferred hyper-parameters, is used as an computationally efficient statistical emulator of the multizone-CFD simulator.
In the post event-detection stage, the Bayesian framework is used to infer the source location and characteristics using the contaminant concentration data obtained through a sensor network.
The Gaussian process emulator of the contaminant fate and transport is used for Markov Chain Monte Carlo sampling to efficiently explore the posterior distribution of source location and characteristics.
Efficacy of the proposed method is demonstrated for a hypothetical contaminant release through multiple sources in a single storey seven room building.
The method is found to infer location and characteristics of the multiple sources accurately.
The posterior distribution obtained using the proposed method is found to agree closely with the posterior distribution obtained by directly coupling the multizone-CFD simulator with the Markov Chain Monte Carlo sampling.
\end{abstract}


\end{frontmatter}
\section{Introduction}
Safety systems in the modern building environments uses sensors that monitor atmospheric parameters and alert in the eventuality of an accident.
With the present day increased threat of use of chemical and biological warfare by terrorist organizations, such a scenario has become a real danger.
Currently, designers are increasingly focussing on development of sensor systems that can detect accidental/deliberate release of hazardous contaminant, and also suggest an appropriate evacuation plan to ensure safety of occupants \cite{chen10}.
Since prolonged exposure of the occupants to the hazardous contaminants may result in serious health conditions including death \cite{zhai03}, rapid source localization by the sensor system is essential.
Considering that majority of individuals are expected to spend upto 90\% of time in an indoor environment, it is imperative to design a sensor system that can detect, characterize and rapidly locate the accidental or deliberate contaminant release.
The system is expected to aid in detection of airborne contaminant, real-time interpretation of the information to characterize and localize the contaminant source, computationally efficient prediction of contaminant dispersion with associated uncertainty quantification, and subsequent evacuation decisions based on the predictions.

\subsection{Background}
The sensor system often uses contaminant fate and transport models to predict the contaminant dispersion that can aid in source localization and characterization.
Multizone, zonal and computational fluid dynamics (CFD) models are used for simulation of indoor airflow and contaminant dispersion patterns \cite{chen08,chen10}.
Owing to ease of implementation and computational efficiency, multizone models are most widely used for predicting the contaminant dispersion and source localization/characterization \cite{zhai03,SreedharanPhD,sreedharan06,sreedharan07,sreedharan11}.
A multizone model represents any building as a network of well-mixed zones connected by flow paths like doors, windows, leaks etc.
The airflow and contaminant transport between the zones is calculated using adjustment of zone pressures that balances mass flow through the paths \cite{SreedharanPhD,nbsir87,feustel99}.
The outdoor environment is modeled as an additional unbounded zone.
Although used widely, limitations of the multizone models, especially related to the well-mixed assumption, are extensively reported in the literature \cite{baughman94,richmond06,SreedharanPhD}.
Zonal models represent intermediate fidelity between multizone and CFD models, wherein large well-mixed zones are further divided into smaller subzones \cite{mora02}.
Zonal models use conservation of mass, conservation of energy and pressure gradients to model airflow and contaminant dispersion \cite{wurtz99}.
Computational fluid dynamics (CFD) models numerically solves governing equations of fluid flow and contaminant dispersions \cite{zhai03}.
The CFD models provide detailed airflow and contaminant distribution inside a room \cite{nielsen04}.
Although most accurate amongst three, computational cost requirement prohibits use of CFD models for rapid source localization and characterization \cite{mora02}.
There are recent research efforts to integrate CFD with multizone models \cite{Tan05,WangPhD,Wang07,Wang10} (termed hereafter as mutlizone-CFD model).
The multizone-CFD modeling approach models one of the zones using CFD, while the resultant solution is coupled with other well-mixed zones using appropriate boundary conditions.
This paper uses the integrated multizone-CFD model for rapid source localization and characterization. 
\subsection{Motivation}
Traditional deterministic approaches for sensor data fusion and interpretation, like optimization \cite{mahar97}, Kalman filtering \cite{federspiel} and backward methods \cite{atmadja01}, are found inappropriate by the researchers in the context of rapid contaminant source localization and characterization \cite{sohn02}.
Owing to the ability to provide the event probability distribution, and associated ease in the uncertainty analysis post event detection, current state of the art for source localization and characterization mainly focusses on probabilistic methods \cite{liu07}.
Liu and Zhai \cite{liu09} have explored adjoint probability method for rapid contaminant source localization.
The method derives adjoint equations for backward probability calculations using the multi-zone contaminant fate and transport model.
Efficacy of the method is demonstrated for contaminant release in a multi-room residential house and a complex institutional building.

Main aim of the present research work is to develop a MCMC-based Bayesian framework that can aid the sensor system to rapidly localize and characterize the contaminant source in case of the event detection.
Main advantage of the Bayesian inference method is that it can admit prior information and estimates complete probability distributions of the uncertain parameters, as against point estimates provided by optimization based methods.
Sohn et al. \cite{sohn02} have proposed a computationally efficient Bayes Monte Carlo method for real-time data interpretation and rapid source localization.
The method is divided in two stages. In first stage, a large database of simulation runs for all the possible scenarios is collected that sufficiently represent uncertainty.
In the second stage, Bayesian updating of the probability for each collected data is obtained after the event detection.
See Sreedharan et al. \cite{SreedharanPhD,sreedharan06,sreedharan07,sreedharan11} for details of recent applications of the Bayes Monte Carlo method.

Though computationally efficient, the Bayes Monte Carlo method essentially is an approximate formulation of the Bayesian inference which can not exploit full capabilities of the Bayesian framework, including ability to handle arbitrary priors and uncertainty in the simulation model.
Rather, if a large number of simulation runs are possible in real time, the MCMC-based Bayesian inference is preferred over the Bayes Monte Carlo method \cite{SreedharanPhD}.
However, currently there is no reported exposition of the MCMC-based Bayesian inference for rapid source localization and characterization in the open literature.

Implementation of the MCMC based Bayesian framework for sensor systems is challenging due to: 1) necessity of rapid real-time inference to ensure successful evacuation with minimum losses; 2) transient nature of the underlying phenomenon; and 3) requirement of large number of MCMC samples (often in the range of $10^3$-$10^6$) for acceptable accuracy.
The problem is further exacerbated by the often large scale nature of the phenomenon being monitored.
Note that items 2) and 3) necessitates large number of dynamic simulator runs, which contradicts with item 1), rendering the MCMC based Bayesian framework intractable for the sensor systems.
This paper proposes computationally efficient Gaussian Process Emulator (GPE) \cite{ken_gpe} based approach for rapid real-time inference in view of dynamic simulators.

\subsection{Proposed Method}
Considering the improved fidelity of the multizone-CFD model over the multizone model, coupled with the accuracy of the MCMC-based Bayesian inference over the Bayes Monte Carlo method, the MCMC-based Bayesian inference using the multizone-CFD model is expected to provide more accurate source localization and characterization as compared to the multizone model based Bayes Monte Carlo method.
However, despite of the significant computational advantage over the CFD implementation, the multizone-CFD model remains computationally prohibitive for MCMC-based rapid source localization and characterization.
This paper proposes a Gaussian Process Emulator (GPE) based Bayesian framework, that can use multizone-CFD model in the context of rapid source localization and characterization.
The proposed approach follows Bayesian inference method of Kennedy and O'Hagan \cite{koh01}, where computer simulator is calibrated using limited number of experimental observations and simulation runs (see also Higdon et al.\cite{HigdonJSC04}, Goldstein and Rougier \cite{Goldstein04}).
The proposed approach treats computer output as a random function \cite{adler}, with the associated probability distribution modeled through a Gaussian process prior.
The Gaussian process prior for representation of uncertain simulator outputs is extensively explored in the literature \cite{sacks89,Oakley02}, with associated hyper-parameters predicted using the maximum likelihood estimates \cite{welch92} or Bayesian inference \cite{CurrinJASA91}.
Conditional on the hyper-parameters and a set of simulator outputs obtained at different input settings, mean of the Gaussian process acts as a computationally efficient statistical emulator of the simulator.
See O'Hagan \cite{ohagan_tut} for detailed tutorial on building the GPE for a simulator, while Kennedy et al. \cite{ken_gpe} may be referred for discussion on some of the case studies.
However, these approaches concern statistical emulation of single-output static simulators.
Conti and O'Hagan\cite{conti10} have extended the GPE method for statistical emulation of dynamic simulators.

This paper adapts the GPE for dynamic simulators proposed by Conti and O'Hagan \cite{conti10} to the multizone-CFD model.
The resultant emulator is used in the Bayesian framework, wherein computational efficiency of the emulator over the simulator is used for rapid source localization and characterization.
The proposed method first uses dynamic simulator output data to derive the GPE, which is then used in the Bayesian framework to infer source location and characteristics using the experimental observations.

The method proposed in this paper advances the current state of the art as follows: a) the method provide MCMC-based Bayesian inference using multizone-CFD model, whereas earlier methods reported in the literature are limited to Bayes Monte Carlo approaches using the multizone models; b) Gaussian process emulator based approach is proposed for efficient Bayesian inference; c) the method provide ability to consider model structural uncertainty, which is not treated in the earlier expositions.

Rest of the paper is organized as follows: detailed problem formulation is presented in section 2.
Section 3 provide details of the emulator for dynamic system simulators.
In section 4, the proposed Bayesian framework for rapid source localization and characterization is discussed in detail.
In section 5, efficacy of the proposed method is demonstrated for a synthetic test case of a hazardous contaminant release in a single storey seven room building.
The paper is summarized and concluded in section 6.


\section{Problem Formulation}
This paper concerns a sudden accidental/deliberate release of contaminant in a building that may cause serious health hazards, including death, to the occupants if exposed over a prolonged period of time.
Although released locally, the contaminant diffuses rapidly through flow paths like doors, windows and leakages, affecting occupants throughout the building.
The building is often equipped with sensors that can detect and measure the amount of contaminant present in a room.
The sensor data is collected over a period of time, which is then used to decide the evacuation strategy and the containment plan, including appropriate air-handling unit actions and source extinguishing strategies.
However, success of the control and evacuation strategy depends on the knowledge of source location and characteristics, which is inferred using the Bayesian framework.
Typically, the source is characterized by specifying the time of activation, $S_t$, and the amount released, $S_a$.
Present paper demonstrates the proposed method for possibly multiple number of sources, $S_N$, while each source is localized by specifying the zone in which the sources are active, $Z$, and xy-coordinate of each source in the zone, $(x_i,y_i)$.
Note that the Bayesian framework relies on ability to accurately predict the contaminant fate and transport for a given source location and characteristics.

\subsection{Integrated Multizone-CFD Model}
Multizone model represents a building using a network of well-mixed zones, each zone often representing a room or compartment connecting to rest of the building through flow paths.
The model account for influences of the internal air flows, which are generated  by pressure differences between the zones.
The multizone model uses internal air flows, coupled with the atmospheric and outdoor wind conditions, to predict contaminant dispersion inside a building.
Wang et al. \cite{WangPhD,Wang07,Wang10} have coupled a multizone model CONTAM \cite{contam} with a zero-turbulence CFD model.
The program define one of the zone as a CFD-zone, where full CFD analysis is used, while the resultant air and contaminant properties are linked with other zones to embed the CFD-zone with CONTAM.
Further, an external coupling is provided to link information on outdoor air pressure and contaminant concentration to indoor building.
This subsection briefly describes the integrated multizone-CFD model.

The multizone model estimates the airflow and the contaminant dispersion between the zones $i$ and $j$, through the flow path $ij$, using the pressure drop across the path $\Delta P_{ij}$.
The model uses a power-law function to calculate the airflow rate, $F_{ij}$, through the flow path $ij$ as \cite{WangPhD}
\begin{equation}
F_{ij} = c_{ij} \left(\frac{P_i - P_j}{\mid P_i - P_j \mid} \right) {\mid \Delta P_{ij} \mid}^{n_{ij}},
\end{equation}
where $c_{ij}$ is flow coefficient, $n_{ij}$ is flow exponent while $P_i$ and $P_j$ are total pressures in zone $i$ and $j$ respectively.
For each zone $j$, the multizone model evaluates steady state air mass balance using
\begin{equation}
\sum_i c_{ij} \left(\frac{P_i - P_j}{\mid P_i - P_j \mid} \right) {\mid \Delta P_{ij} \mid}^{n_{ij}} + F_j = 0,
\end{equation}
where $F_j$ is the air mass source in the zone $j$.
Contaminant steady state mass balance for a species $\alpha$ is similarly obtained by
\begin{equation}
\sum_{i} F_{ij} C_\alpha + S_j = 0,
\label{cont_massb}
\end{equation}
where $S_j$ is the contaminant source in the zone $j$, while $C_\alpha$ is a contaminant concentration defined such that
\begin{equation}
C_\alpha = \begin{cases} C_{\alpha_i}, & \mbox{if airflow is from zone $i$ to $j$}  \\ C_{\alpha_j}, & \mbox{if airflow is from zone $j$ to $i$} \end{cases};
\end{equation}
$C_{\alpha_i}$ and $C_{\alpha_j}$ are the contaminant concentrations in zone $i$ and $j$ respectively.

The CFD model solves a set of partial differential governing equations for conservation of mass, momentum and energy inside the CFD zone.
The governing equations for steady state flow are given by
\begin{equation}
\nabla(\rho \boldsymbol{V} u) - \Gamma_u \nabla^2 u = S_u,
\label{cfd}
\end{equation}
where $u$ is a variable of conservation equations, $\rho$ is density, $\boldsymbol{V}$ is velocity vector, $\Gamma_u$ is diffusion coefficient, and $S_u$ is source.
At each time step, CFD model solves steady state conservation equations \ref{cfd}.

Let the CFD zone, $c$, be connected to a zone, $i$, using a flow-path $ic$.
For each grid point of the discretized flow path $ic$, CFD model calculates mass flow rate normal to the cell $p$, $f_p$, by
\begin{equation}
f_p = c_{L,p} (P_i + d_{ic} - P_p),
\end{equation}
where $c_{L,p}$ is a linear flow coefficient, $P_i$ is pressure in zone $i$, $d_{ic}$ is a pressure difference between zones $i$ and $c$, while $P_p$ is pressure at a grid point $p$.
Thus, the total mass flow through flow path $ic$ predicted by the CFD model is given by
\begin{equation}
F^C_{ic} = \sum^{n_g}_{p=1} f_p,
\end{equation}
where $n_g$ is total number of grid points for the flow path.
The multizone model predicts the total mass flow through flow path $ic$ as
\begin{equation}
F^M_{ic} = c_{L,ic} (P_i + d_{ic} - P_{d,ic}),
\end{equation}
where $P_{d,ic}$ is the average downwind total pressure for path $ic$.
Thus, the coupling between CFD and multizone models is obtained by ensuring
\begin{equation}
\sum_{k} \left(\mid F^M_{ic} - F^C_{ic} \mid \right)_k \leq \epsilon
\end{equation}
for all connecting flow paths $k$, where $\epsilon$ is a convergence criterion.
Using the total mass flow, contaminant concentration in each zone is estimated using Eq. (\ref{cont_massb}).

\begin{figure}[h!]
  \begin{center}
   \includegraphics[width=3.0in, height=2.65in] {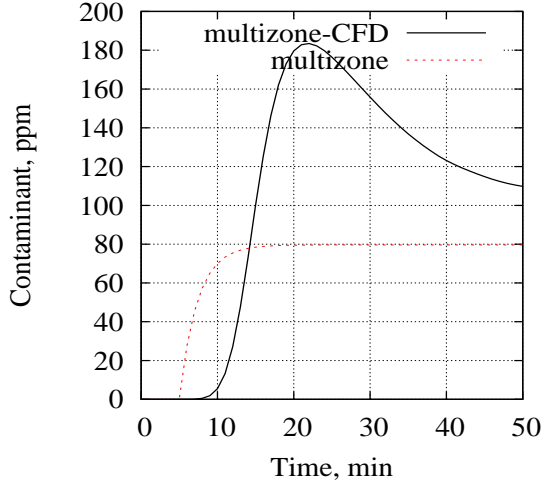}
  \end{center}
 \caption{Comparison of prediction by multizone and multizone-CFD model}
\label{comp_cfd}
\end{figure}

In the present paper, the coupled multizone-CFD model available with CONTAM \cite{contamSite,WangPhD,Wang07,Wang10,contam} is used to simulate the contaminant fate and transport.
The room containing active contaminant sources is always defined as a CFD-zone, while other rooms are simulated using multizone model.
Transient contaminant concentration in each zone is output of the multizone-CFD model.
To motivate the choice of multizone-CFD model over the multizone model, it is imperative to investigate the difference between transient contaminant concentration predictions, as shown in Figure \ref{comp_cfd}.
From the figure, significant difference between predictions can be observed, which may result in erroneous localization and characterization of contaminant sources.
The main motivation for the present research work is to develop a Bayesian inference method that can use the more accurate multizone-CFD model for rapid localization of contaminant source in an indoor environment.

\subsection{Bayesian Framework}
This subsection presents reformulation of the rapid source localization and characterization problem in the Bayesian inference terminology.
For notational convenience and brevity, the formulation is presented for a single contaminant species, however, the method can be extended without any change for multiple species.
Let $y_j(t)=T\left(\boldsymbol{x},\boldsymbol{\theta};t \right)$, represent the multizone-CFD model, where $\boldsymbol{x}\in \mathcal{\boldsymbol{X}}$ is a set of deterministic inputs, $\boldsymbol{\theta}\in {\boldsymbol{\Theta}}$ is a set of uncertain parameters, while $y_j(t) = \{C_j(t)\}$ is a contaminant concentration in the $j^{th}$ zone at time $t$.
For the multizone-CFD model, $\boldsymbol{x}$ typically  consists of building description including rooms and flow path specifications, air-handling unit, atmospheric and wind conditions, etc., while, the uncertain parameters are $\boldsymbol{\theta}=[S_N, Z, S_a, S_t, \{(x_i,y_i);i=1,..,S_N\}]$.
For further notational convenience, define $\boldsymbol{y}_j = \{y_j(t); t \in \mathcal{R}^+\}$ as a function representing the transient contaminant concentration, such that
\begin{equation}
\boldsymbol{y}_j = T\left(\boldsymbol{x},\boldsymbol{\theta} \right).
\label{funct_op}
\end{equation}
Note that Eq. (\ref{funct_op}) represents a simulator with function as output, thus, explicit dependence on $t$ is removed.

Let $\boldsymbol{\hat{\theta}}$ be the set of `true' but unknown source location and characteristics, that need to be inferred for future decisions, including control and evacuation strategies.
To account for possible deficiencies of the multizone-CFD model, the simulator output is assumed to deviate from the `true' system response even on specification of $\boldsymbol{\hat{\theta}}$.
In the present paper, this deviation is modeled as \cite{koh01}
\begin{equation}
\boldsymbol{\zeta}_j = T\left(\boldsymbol{x},\boldsymbol{\hat{\theta}} \right) + \boldsymbol{\delta}_j,
\label{zeta1}
\end{equation}
where $\boldsymbol{\zeta}_j = \{\zeta_{j}(t);~ t \in R^+ \}$ is the `true' system response, while $\boldsymbol{\delta}_j=\{\delta_j(t);~ t \in R^+ \}$ is known as a discrepancy function.

Let the building be equipped with sensors in $M$ zones that detect and measure the contaminant concentration.
Let the sensor data be collected for $N$ discrete time instances.
The relationship between the sensor measurement  and the `true' contaminant concentration for $j^{th}$ zone at $i^{th}$ time instance is given by
\begin{equation}
{y_e}_{j}(t_i) = \zeta_{j}(t_i) + \epsilon_j(t_i),
\label{ExpTrue}
\end{equation}
where $\epsilon_j\left(t_i\right)$ denotes the sensor measurement uncertainty.
For notational convenience, define a set of sensor observations $\boldsymbol{Y_e} = \{{y_e}_{j}(t_i); i=1,...,N; j=1,...,M \}$.
Similarly define $\boldsymbol{\delta} = \{\delta_{j}(t_i); i=1,...,N; j=1,...,M \}$.
Using $\boldsymbol{Y_e}$, $\boldsymbol{\hat{\theta}}$ and $\boldsymbol{\delta}$ can be inferred through the Bayes theorem as
\begin{equation}
p(\boldsymbol{\hat{\theta}},\boldsymbol{\delta} \mid \boldsymbol{Y_e}) \propto p(\boldsymbol{Y_e} \mid \boldsymbol{\hat{\theta}},\boldsymbol{\delta}) \times p(\boldsymbol{\hat{\theta}},\boldsymbol{\delta}),
\label{BayesTh}
\end{equation}
where $p(\boldsymbol{\hat{\theta}},\boldsymbol{\delta})$ is the prior, $p(\boldsymbol{Y_e} \mid \boldsymbol{\hat{\theta}},\boldsymbol{\delta})$ is the likelihood, and $p(\boldsymbol{\hat{\theta}},\boldsymbol{\delta} \mid \boldsymbol{Y_e})$ is the posterior probability distribution.

In the present paper, $\epsilon_j\left(t_i \right)$ is assumed to be a zero-mean normally distributed random variable with covariance function
\begin{equation}
\Sigma_{e_j} = \sigma^2_{e_j} I_N,
\end{equation}
where $\sigma_{e_j}$ is the standard deviation of uncertain experimental observations, while $I_N$ is the $N \times N$ identity matrix.
The prior uncertainty in $\boldsymbol{\delta_j}$ is specified using a zero-mean Gaussian process with covariance function
\begin{equation}
\Sigma_{\delta_j}(t_1,t_2) = \sigma^2_{\delta_j} \exp \left(-\lambda_j (t_1-t_2)^2 \right),
\label{CovDisc}
\end{equation}
where $\sigma^2_{\delta_j}$ and $\lambda_j$ are uncertain hyper-parameters.
In the full Bayesian analysis, $\sigma^2_{\delta_j}$ and $\lambda_j$ are also inferred using the Bayes theorem.
Since the method presented in this paper concerns inference of parameters for decisions involving control/evacuation strategies, $\sigma^2_{\delta_j}$ and $\lambda_j$ are assumed to be fixed\footnote{Note that full Bayesian analysis can be used a-priory to infer the hyper-parameters $\sigma^2_{\delta_j}$ and $\lambda_j$.}.
Using the probability distribution of $\boldsymbol{\epsilon}$ and marginalization of $\boldsymbol{\delta}$, the posterior probability distribution is given by
\begin{equation}
p(\boldsymbol{\hat{\theta}},\mid \boldsymbol{Y_{e_j}}, \boldsymbol{\sigma^2_\delta},\boldsymbol{\lambda}) \propto \mid \Sigma_j \mid^{-\frac{1}{2}} \prod^{M}_{j=1} \exp\left( -\frac{1}{2} \boldsymbol{d_j}^T \Sigma^{-1}_j \boldsymbol{d_j} \right) \times p(\boldsymbol{\hat{\theta}}),
\label{BayFin}
\end{equation}
where $\boldsymbol{d_j}=\{y_{e_{j}}(t_i) -T(\boldsymbol{x},\hat{\boldsymbol{\theta}}; t_i); i=1,...,N; j=1,...,M\}$, $\boldsymbol{\sigma^2_\delta}=\{\sigma^2_{\delta_j}; j=1,...,M \}$, $\boldsymbol{\lambda}=\{\lambda_j; j=1,...,M\}$ and $\Sigma_j=\Sigma_{\delta_j} + \Sigma_{e_j}$.
Solution of Eq. (\ref{BayFin}) require sampling from the posterior distribution using Markov Chain Monte Carlo (MCMC) method.
MCMC method can be implemented to sample from the probability distribution of a random vector $\boldsymbol{\phi}$, $p(\boldsymbol{\phi})$, using the Metropolis-Hastings algorithm as follows \cite{metr, HastingsBio70}:
\begin{algorithm}
\caption{Metropolis-Hastings Algorithm for MCMC Sampling}
\label{alg_metr}
\begin{algorithmic}[1]
 \STATE Initialize the chain at $\boldsymbol{\phi} = \boldsymbol{\phi}_0$
 \FOR{$i=1$ \TO total\_no\_samples}
  \STATE Sample a trial point $\boldsymbol{\phi}_*$ from proposal distribution $f(\boldsymbol{\phi}_*\mid \boldsymbol{\phi}_{i-1})$
  \STATE Calculate acceptance probability
\begin{equation}
A\left(\boldsymbol{\phi}_*,\boldsymbol{\phi}_{i-1}\right) = \min \left\{1, \frac{p(\boldsymbol{\phi}_*) f(\boldsymbol{\phi}_*\mid \boldsymbol{\phi}_{i-1})}{p(\boldsymbol{\phi}_{i-1}) f(\boldsymbol{\phi}_{i-1} \mid \boldsymbol{\phi}_* )} \right\}
\end{equation}
 \STATE Generate a uniform random variable $\mathcal{U}$
  \IF{$\mathcal{U} < A\left(\boldsymbol{\phi}_*,\boldsymbol{\phi}_{i-1}\right)$}
   \STATE $\boldsymbol{\phi}_{i} = \boldsymbol{\phi}_*$
  \ELSE
   \STATE $\boldsymbol{\phi}_{i} = \boldsymbol{\phi}_{i-1}$
 \ENDIF
\ENDFOR
\end{algorithmic}
\end{algorithm}

Note that the implementation of the MCMC require solution of $T(\boldsymbol{x},\cdot)$ for each sample, rendering the Bayesian framework intractable for computationally expensive simulators.
The method proposed in this paper uses Gaussian process emulator (GPE) of the simulator in the MCMC sampling for rapid real time inference.
Following section provide details of building a GPE for the dynamic simulator $T\left(\boldsymbol{x},\cdot \right)$.

\section{Gaussian Process Emulator for Dynamic Simulator}
For a given $\boldsymbol{x}$, the simulator $T\left(\boldsymbol{x},\boldsymbol{\theta}\right)$ maps a $d$-dimensional input $\boldsymbol{\theta}\in \boldsymbol{\Theta} \subset \mathcal{R}^{d}$ to a transient output $\boldsymbol{y}\in \mathcal{T}\subset \mathcal{R}^\mathcal{R}$, where $\boldsymbol{y}$ is a function of continuous time.     
An emulator is built using a subset of transient response $\boldsymbol{\hat{y}}\subset \boldsymbol{y}$, which is treated as a $q$-variate output of the simulator, thus $\boldsymbol{\hat{y}} \in \mathcal{R}^{q}$.
The simulator is deterministic in a sense that repeated simulation runs at a given input setting always returns same output.
However, the simulator output is considered uncertain as the simulator runs at all the possible values of $\boldsymbol{\theta}$ can not be obtained for computationally intensive simulators.
Thus, $T(\boldsymbol{x},\cdot)$ is treated as a random function,\footnote{For a function with univariate output, a random function can be considered as a sample from a stochastic process $\mathcal{F}: \mathcal{X} \longrightarrow \mathcal{R}^{\mathcal{R}}$.
A $q$-variate random function is a generalization $\mathcal{F}: \mathcal{X} \longrightarrow ({\mathcal{R}^q})^{\mathcal{R}}$. See Adler \cite{adler} for details.} with a probability distribution quantified using a Gaussian process \cite{sacks89,CurrinJASA91,Rasmussen,koh01,ohagan_tut}.
Following Conti and O'Hagan \cite{conti10}, uncertainty in the random function is specified using a $q$-dimensional Gaussian process as
\begin{equation}
T(\boldsymbol{x},\cdot) \sim \mathcal{N}_q\left(\mathbf{m}(\cdot),c(\cdot,\cdot)\Sigma \right),
\end{equation}
where $\mathbf{m}(\cdot)$ is mean and $c(\cdot,\cdot)\Sigma$ is a covariance structure of the Gaussian process.
Often, the mean is modeled as
\begin{equation}
\mathbf{m}(\cdot) = \mathcal{B}^T \mathbf{h}(\cdot),
\end{equation}
where $\mathbf{h}(\cdot)=\left[h_1(\cdot), h_2(\cdot),....,h_m(\cdot) \right]^T$ is a vector of $m$ regression functions, while $\mathcal{B}\in \mathcal{R}^{m \times q}$ is a matrix of regression coefficients with each column given by $\boldsymbol{\beta}=\left[\beta_1, \beta_2, ...., \beta_m \right]^T$.
Though an arbitrary regression model can be used, literature suggests a linear model suffice for majority of the applications \cite{koh01}, thus $\mathbf{h}=\left[1,\boldsymbol{\theta}\right]^T$ and $m=d+1$.

Covariance function of the Gaussian process is given by
\begin{equation}
cov\left(T(\mathbf{x},\boldsymbol{\theta}_1), T(\mathbf{x},\boldsymbol{\theta}_2)\right) = c(\boldsymbol{\theta}_1,\boldsymbol{\theta}_2) \Sigma,
\end{equation}
where $c(\boldsymbol{\theta}_1,\boldsymbol{\theta}_2)$ is a positive-definite correlation function, while $\Sigma \in \mathcal{R}^{q \times q}_{+}$ is a $q\times q$ positive definite matrix.
In the present work, a square exponential correlation function is used
\begin{equation}
c(\boldsymbol{\theta}_1,\boldsymbol{\theta}_2) = \exp \left(-(\boldsymbol{\theta}_1-\boldsymbol{\theta}_2)^T \Lambda (\boldsymbol{\theta}_1 - \boldsymbol{\theta}_2) \right),
\end{equation}
where $\Lambda$ is a diagonal matrix with diagonal elements given by a vector of $d$ correlation length parameters $\boldsymbol{\lambda}$.
Parameters $\mathcal{B}$, $\Sigma$ and $\boldsymbol{\lambda}$ are treated as uncertain hyper-parameters.
Weak non-informative prior is used for $\mathcal{B}$ and $\Sigma$,
\begin{equation}
p(\mathcal{B},\Sigma \mid \boldsymbol{\lambda}) \propto {\mid \Sigma \mid}^{-\frac{q+1}{2}},
\label{prior}
\end{equation}
while prior for $\boldsymbol{\lambda}$ is left unspecified.

A set of $n$ simulation runs at design points $\mathcal{\mathbf{S}}=\left[\boldsymbol{\theta}_1,\boldsymbol{\theta}_2,...., \boldsymbol{\theta}_n \right] \subset \boldsymbol{\Theta}$ is used to build an emulator.
Let $\boldsymbol{D} \in \mathcal{R}^{n\times q}$ define a $n\times q$ matrix of simulator outputs.
From the Bayesian perspective, an emulator is defined as posterior distribution of the random function $T(\boldsymbol{x},\cdot)$ given a set of simulation runs $\boldsymbol{D}$.
Conditional on hyper-parameters  $\mathcal{B}$, $\Sigma$ and $\boldsymbol{\lambda}$, probability distribution of $\boldsymbol{D}$ is given by \cite{conti10}
\begin{equation}
p(\boldsymbol{D} \mid \mathcal{B}, \Sigma, \boldsymbol{\lambda}) \sim \mathcal{N}_{n,q}\left(\mathcal{H}\mathcal{B},\mathcal{A}\Sigma\right),
\label{probD}
\end{equation}
where $\mathcal{H}^T=\left[\mathbf{h}(\boldsymbol{\theta}_1),\mathbf{h}(\boldsymbol{\theta}_2),...,\mathbf{h}(\boldsymbol{\theta}_n)\right] \in \mathcal{R}^{m\times n}$ and $\mathcal{A}=c(\boldsymbol{\theta}_i,\boldsymbol{\theta}_j) \in \mathcal{R}^{n\times n}$ is a correlation matrix for a design set $\mathcal{\boldsymbol{S}}$.
Using Eq. (\ref{probD}) as likelihood and prior given by Eq. (\ref{prior}), posterior distribution of hyper-parameters is given by
\begin{equation}
p(\mathcal{B}, \Sigma, \boldsymbol{\lambda}\mid \boldsymbol{D}) \propto  \mathcal{N}_{n,q}\left(\mathcal{H}\mathcal{B},\mathcal{A}\Sigma\right) {\mid \Sigma \mid}^{-\frac{q+1}{2}} p(\boldsymbol{\lambda}).
\label{post_hyper}
\end{equation}

Conditional on the posterior distribution of hyper-parameters and $\boldsymbol{D}$, the emulator is defined as \cite{conti10}
\begin{equation}
p(T(\boldsymbol{x},\cdot)\mid \mathcal{B}, \Sigma, \boldsymbol{\lambda}, \boldsymbol{D}) \sim \mathcal{N}_q(\mathbf{m}^{*}(\cdot),c^{*}(\cdot,\cdot)\Sigma),
\label{emul1}
\end{equation}
where
\begin{equation}
\begin{split}
& \mathbf{m}^{*}(\boldsymbol{\theta}) = \mathcal{B}^T \mathbf{h}(\boldsymbol{\theta}) + (\boldsymbol{D} - \mathcal{H} \mathcal{B})^T \mathcal{A}^{-1} \mathbf{r}(\boldsymbol{\theta}) \\
& c^{*}(\boldsymbol{\theta}_1,\boldsymbol{\theta}_2) = c(\boldsymbol{\theta}_1,\boldsymbol{\theta}_2) - \mathbf{r}^T(\boldsymbol{\theta}_1) \mathcal{A}^{-1} \mathbf{r}(\boldsymbol{\theta}_2),
\end{split}
\label{emul1_mc}
\end{equation}
while $\mathbf{r}^T(\cdot) = \left[c(\cdot,\boldsymbol{\theta}_1),....,c(\cdot,\boldsymbol{\theta}_n)\right]\in \mathcal{R}^n$.
Equation (\ref{emul1}) is a statistical emulator of the simulator \cite{sacks89,welch92,ken_gpe,conti10,ohagan_tut}, with mean and covariance (Eq. (\ref{emul1_mc})) acting as interpolator and associated expected error, respectively.

Note that Eq. (\ref{emul1_mc}) requires sampling from posterior distribution of hyper-parameters, imposing significant computational cost.
Thus, if the analytical form is available,  marginalization of hyper-parameters can render implementation of the statistical emulator computationally tractable \cite{conti10}.
First, marginalization of $\mathcal{B}$ from Eqs. (\ref{prior})-(\ref{emul1}) gives
\begin{equation}
p(T(\boldsymbol{x},\cdot)\mid \Sigma, \boldsymbol{\lambda}, \boldsymbol{D}) \sim \mathcal{N}_q(\mathbf{m}^{**}(\cdot),c^{**}(\cdot,\cdot)\Sigma)
\label{emul2}
\end{equation}
where
\begin{equation}
\begin{split}
& \mathbf{m}^{**}(\boldsymbol{\theta}) = \mathcal{\hat{B}}^T \mathbf{h}(\boldsymbol{\theta}) + (\boldsymbol{D} - \mathcal{H} \mathcal{\hat{B}})^T \mathcal{A}^{-1} \mathbf{r}(\boldsymbol{\theta}) \\
& c^{**}(\boldsymbol{\theta}_1,\boldsymbol{\theta}_2) = c^{*}(\boldsymbol{\theta}_1,\boldsymbol{\theta}_2) + [\mathbf{h}(\boldsymbol{\theta}_1)- \mathcal{H}^T \mathcal{A}^{-1} \mathbf{r}(\boldsymbol{\theta}_1)]^T \\
& \qquad \qquad (\mathcal{H}^T \mathcal{A}^{-1} \mathcal{H})^{-1} [\mathbf{h}(\boldsymbol{\theta}_2)- \mathcal{H}^T \mathcal{A}^{-1} \mathbf{r}(\boldsymbol{\theta}_2)].
\end{split}
\label{emul_fin}
\end{equation}
Here, $\mathcal{\hat{B}}$ is a generalized least square estimate of $\mathcal{B}$ given by
\begin{equation}
\mathcal{\hat{B}} = (\mathcal{H}^T \mathcal{A}^{-1} \mathcal{H})^{-1} \mathcal{H}^T \mathcal{A}^{-1} \boldsymbol{D}.
\label{bgls}
\end{equation}
Further integrating out $\Sigma$ from Eq. (\ref{emul2}) to obtain \cite{conti10}
\begin{equation}
p(T(\boldsymbol{x},\cdot)\mid \boldsymbol{\lambda}, \boldsymbol{D}) \sim \mathcal{T}_q(\mathbf{m}^{**}(\cdot),c^{**}(\cdot,\cdot)\hat{\Sigma};n-m),
\label{emul3}
\end{equation}
where $\mathcal{T}_q$ is a Student's T process, while $\hat{\Sigma}$ is generalized least square estimator of $\Sigma$, which is given by
\begin{equation}
\hat{\Sigma} = \frac{(\boldsymbol{D} - \mathcal{H} \mathcal{\hat{B}})^T \mathcal{A}^{-1} (\boldsymbol{D} - \mathcal{H} \mathcal{\hat{B}})}{n-m}.
\label{sgls}
\end{equation}

Final step to build an emulator involves marginalization of $\boldsymbol{\lambda}$, however, analytical solution for the resultant integration is not available and requires use of sampling techniques.
Instead, the literature suggests fixing the values of correlation length parameters using Maximum Posteriory Estimate (MPE), Eq. (\ref{emul3}) conditional on MPE of $\boldsymbol{\lambda}$ being an emulator of $T(\boldsymbol{x},\cdot)$.

Posterior distribution of $\boldsymbol{\lambda}$ is obtained by marginalization of $\mathcal{B}$ and $\Sigma$ from Eq. (\ref{post_hyper}), which gives
\begin{equation}
p(\boldsymbol{\lambda}\mid \boldsymbol{D}) \propto {\mid \mathcal{A} \mid}^{-q/2} {\mid \mathcal{H}^T \mathcal{A}^{-1} \mathcal{H}\mid}^{-q/2} {\mid \boldsymbol{D}^T \mathcal{G} \boldsymbol{D}\mid}^{-(n-m)/2},
\label{postLambda}
\end{equation}
where
\begin{equation}
\mathcal{G} = \mathcal{A}^{-1} - \mathcal{A}^{-1} \mathcal{H} (\mathcal{H}^T \mathcal{A}^{-1} \mathcal{H})^{-1} \mathcal{H}^T \mathcal{A}^{-1}.
\end{equation}
The MPE of $\boldsymbol{\lambda}$ is obtained by maximizing Eq. (\ref{postLambda}) with respect to $\boldsymbol{\lambda}$.
For the emulator, the mean $\mathbf{m}^{**}(\cdot)$ works as an interpolator providing predictions at an unsampled $\boldsymbol{\theta}$, while $c^{**}(\cdot,\cdot)$ provide estimate of uncertainty in the predictions.
Thus, the Gaussian process emulator for the dynamic simulator can be implemented using the following algorithm:
\begin{algorithm}
\caption{Gaussian Process Emulator for the Dynamic Simulator}
\label{alg_gpe_dynsim}
\begin{algorithmic}[1]
 \STATE Select $n$ and a set of design points $\mathcal{\mathbf{S}}=\left[\boldsymbol{\theta}_1,\boldsymbol{\theta}_2,...., \boldsymbol{\theta}_n \right]$ using design of experiments
 \STATE Select $q$ temporal locations
 \FOR{$i=1$ \TO $n$}
  \STATE Simulate $\boldsymbol{y} = T(\boldsymbol{x},\boldsymbol{\theta}_i)$
  \STATE Define $\boldsymbol{D}$ with $i^{th}$ row given by $\boldsymbol{D}_{i}=\{{y}(t_j); j=1,...,q\}$, where $y(t_j)$ represents the simulator output, $T(\boldsymbol{x},\boldsymbol{\theta}_i; t_j)$, at the time instance $t_j$.   
\ENDFOR
\STATE Estimate GLS $\hat{\mathcal{B}}$ using Eq. (\ref{bgls})
\STATE Estimate GLS of $\hat{\Sigma}$ using Eq. (\ref{sgls})
\STATE Estimate MPE of $\boldsymbol{\lambda}$ by maximizing Eq. (\ref{postLambda}) with respect to $\boldsymbol{\lambda}$
\STATE Using the estimates of $\hat{\mathcal{B}}$, $\hat{\Sigma}$ and $\boldsymbol{\lambda}$, the emulator is defined by
\begin{equation*}
\begin{split}
& \mathbf{m}^{**}(\boldsymbol{\theta}) = \mathcal{\hat{B}}^T \mathbf{h}(\boldsymbol{\theta}) + (\boldsymbol{D} - \mathcal{H} \mathcal{\hat{B}})^T \mathcal{A}^{-1} \mathbf{r}(\boldsymbol{\theta}) \\
\end{split}
\end{equation*}
\end{algorithmic}
\end{algorithm}
\section{Proposed Method}
\subsection{Gaussian Process Emulator for multizone-CFD Simulator}
In the present paper, efficacy of the proposed method is demonstrated for localization and characterization of multiple sources in a building.
Since the current version of coupled multizone-CFD simulator allows only one zone as CFD-zone, the method assumes all the sources be active in a single zone.
For a given number of active sources in the zone, multizone-CFD simulator provides averaged transient contaminant concentration in each zone.
Thus, each transient response is indexed by number of active sources ($a$), the zone in which sources are active ($b$), and the zone in which contaminant concentration is measured ($c$).
Separate GPEs are built for each combination of ($a,b,c$).

\begin{figure}[h!]
  \begin{center}
   \includegraphics[width=3.0in, height=2.85in] {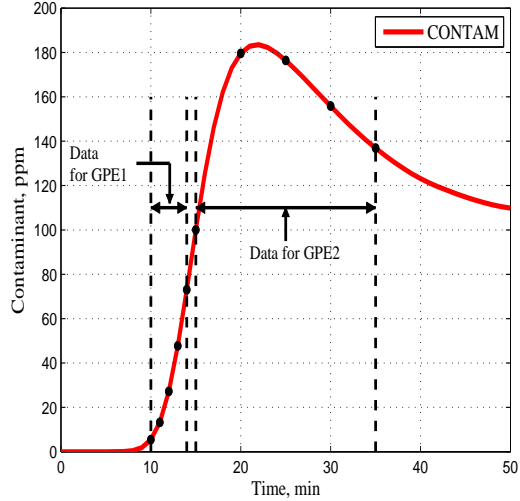}
  \end{center}
 \caption{Result of a CONTAM simulation run}
\label{contam_resp}
\end{figure}

An initial design set, $\boldsymbol{S}_{ini}=\{\boldsymbol{\theta}_i;i=1,...,n_{ini}\}$, is selected using Latin hypercube sampling \cite{McKay79,Munholland96,HeltonSAND0417} and transient simulator responses are obtained for each design point.
A typical response of the simulator is shown in Figure \ref{contam_resp}.
Each transient is divided into two parts, first part consists of $q_1$ closely spaced data points collected just after the source activation, while the second part consists of much more coarsely spaced $q_2$ points.
A set of $n_{ini} \times q_1$ data points, $\boldsymbol{D}^{ini}_1$, is defined using transients obtained at $\boldsymbol{S}_{ini}$.
Conditional on $\boldsymbol{D}^{ini}_1$, MPE estimate of $\boldsymbol{\lambda}$ are obtained by maximizing Eq. (\ref{postLambda}).
Conditional on the MPE estimate of $\boldsymbol{\lambda}$, an additional set of design points $\boldsymbol{\theta}^{new}$ is selected as
\begin{equation}
\arg \max_{\theta \in \Theta} c^{**}(\boldsymbol{\theta}, \boldsymbol{\theta}).
\end{equation}
The additional set of design points is generated sequentially till the maxima of $c^{**}(\cdot,\cdot)$ is below certain pre-defined value.
It may be noted that during the process of selecting additional design points, $\boldsymbol{\lambda}$ is kept constant, while generalized least square estimates $\hat{\mathcal{B}}$ and $\hat{\Sigma}$ are calculated after addition of each new design point.
For this enhanced design set $\boldsymbol{S}$, set of $n \times q_1$ data points, $\boldsymbol{D}_1$, and $n \times q_2$ data points, $\boldsymbol{D}_2$, are defined.
Conditional on $\boldsymbol{\lambda}$, generalized least square estimates $\hat{\mathcal{B}_1}$ and $\hat{\Sigma_1}$ are calculated using $\boldsymbol{D}_1$.
Using the same value of $\boldsymbol{\lambda}$, estimates of $\hat{\mathcal{B}_2}$ and $\hat{\Sigma_2}$ are similarly calculated using $\boldsymbol{D}_2$\footnote{Note that the sensor system is expected to detect the contaminant soon after the source activation, thus, the densely spaced points are used for source localization. The coarse $q_2$ points can then be used for predicting the long term fate and transport of the contaminant.}.

\subsubsection{Reconstruction of Transient Contaminant Concentration}
Let $\mathcal{T}_1=\{t_i; i=1,...,q_1\}$ and $\mathcal{T}_2=\{t_{q_1+i}; i=1,...,q_2\}$ be the time instances at which data sets $\boldsymbol{D}_1$ and $\boldsymbol{D}_2$ are defined, respectively.
For an arbitrary $\boldsymbol{\theta}$, let $\boldsymbol{\mu}_1(\boldsymbol{\theta})=\{m^{**}_{1,i}(\boldsymbol{\theta}); i=1,...,q_1\}$ and $\boldsymbol{\mu}_2(\boldsymbol{\theta})=\{m^{**}_{2,i}(\boldsymbol{\theta}); i=1,...,q_2\}$ define the predicted contaminant concentration obtained using GPEs at time instances $\mathcal{T}_1$ and $\mathcal{T}_2$ respectively.
Further, define a vector, $\boldsymbol{\mu}(\boldsymbol{\theta})=\{\boldsymbol{\mu}_1(\boldsymbol{\theta}),\boldsymbol{\mu}_2(\boldsymbol{\theta})\}$, and a matrix
\begin{equation}
\mathcal{A} = \begin{bmatrix} \hat{\Sigma}_1 & \boldsymbol{0} \\ \boldsymbol{0}^T & \hat{\Sigma}_2 \end{bmatrix} ,
\label{sigma_nq}
\end{equation}
where $\boldsymbol{0}$ is a $q_1 \times q_2$ matrix of zeroes.
Conditional on $\boldsymbol{\mu}(\boldsymbol{\theta})$ and $\mathcal{A}$, the contaminant concentration at any time $t$ is given by
\begin{equation}
y(t;\boldsymbol{\theta}) \sim \mathcal{N}(\mu_*,\nu_*).
\label{cont_conc}
\end{equation}
Using multivariate normal theory, mean and variance of the normal distribution (\ref{cont_conc}) are given by
\begin{equation}
\begin{split}
& \mu_* = \boldsymbol{r}_*(t) \mathcal{A}^{-1} \boldsymbol{\mu} \\
& \nu_* = r_{**} - \boldsymbol{r}_*(t) \mathcal{A}^{-1} \boldsymbol{r}^T_*(t)
\end{split}
\label{predTime}
\end{equation}
where, $ \boldsymbol{r}_*(t)=\{cov(t,t_i); i=1,q_1 + q_2\}$ and $r_{**}=cov(t,t)$.
Equation (\ref{predTime}) is used as an emulator to predict long term fate and transport of the contaminant.
The overall procedure for building the proposed GPE is summarized in Algorithm \ref{alg1}.
\begin{algorithm}
\caption{GPE for Multizone-CFD}
\label{alg1}
\begin{algorithmic}[1]
 \STATE Select $\boldsymbol{S}_{ini}=\{\boldsymbol{\theta}_i;i=1,...,n_{ini}\}$ using Latin hypercube sampling
 \STATE Run multizone-CFD for each $\boldsymbol{\theta}_i\in \boldsymbol{S}_{ini}$
 \STATE Using transient response at $q_1$ time instances, create $\boldsymbol{D}^{ini}_1$
 \STATE Estimate $\boldsymbol{\lambda}$ by maximizing Eq. (\ref{postLambda}) conditional on $\boldsymbol{D}^{ini}_1$
 \WHILE{$c^{**}(\cdot,\cdot) \geq $ tolerance}    
 \STATE Conditional on $\boldsymbol{\lambda}$ and $\boldsymbol{D}^{ini}_1$, 
\begin{equation}
\arg \max_{\theta \in \Theta} c^{**}(\boldsymbol{\theta}, \boldsymbol{\theta}).
\end{equation}
 \STATE $\boldsymbol{S}_{ini} = \boldsymbol{S}_{ini} \cup \boldsymbol{\theta}^{new}$
 \STATE Create $\boldsymbol{D}^{ini}_1$ using $\boldsymbol{S}_{ini}$
 \ENDWHILE
 \STATE $\boldsymbol{S} = \boldsymbol{S}_{ini}$ and $\boldsymbol{D}_1 = \boldsymbol{D}^{ini}_1$
 \STATE Create $\boldsymbol{D}_2$ using transient response at $q_2$ time instances for all $\boldsymbol{\theta}_i \in \boldsymbol{S}$
 \STATE Conditional on $\boldsymbol{\lambda}$ and $\boldsymbol{D}_1$, calculate $\hat{\mathcal{B}_1}$ and $\hat{\Sigma_1}$
 \STATE Conditional on $\boldsymbol{\lambda}$ and $\boldsymbol{D}_2$, calculate $\hat{\mathcal{B}_2}$ and $\hat{\Sigma_2}$
 \STATE Use $\boldsymbol{\lambda}$, $\boldsymbol{D}_1$, $\boldsymbol{D}_2$, $\hat{\mathcal{B}_1}$, $\hat{\mathcal{B}_2}$, $\hat{\Sigma_1}$ and $\hat{\Sigma_2}$ to predict long term transient contaminant concentration.
\end{algorithmic}
\end{algorithm}
\subsection{Rapid Source Localization and Characterization}
Consider a building with total $N_z$ zones, with $N_s$ maximum possible active sources in each zone.
For each possible combination of $a\in N_s$, $b\in N_z$ and $c\in N_z$ the emulator $\mathcal{E}_{a,b,c}\left(\boldsymbol{x},\cdot\right)$ is built using Algorithm \ref{alg1}.
The proposed GPE is used in the Bayesian framework for rapid source localization and characterization in the indoor building environment.

In the present paper, the proposed method is demonstrated for maximum possible 3 sources in a zone.
The prior uncertainty in number of sources, $S_n$, is given by
\begin{equation}
p(S_n) = \frac{1}{N_s}.
\label{priorNs}
\end{equation}
Location of each source is assumed to be completely unknown with prior given by uniform distribution.
Thus,
\begin{equation}
\begin{aligned}
& p({x_i},{y_i},Z) & = &  p(x_i,y_i\mid Z) \times p(Z) \\
& ~ & = & \frac{1}{A_z} \times \frac{1}{N_z},
\end{aligned}
\label{priorloc}
\end{equation}
where $A_z$ is area of zone $Z$.
$S_a$ and $S_t$ are assumed to be completely unknown with the range of possible values as only available information.
Let $S_a\in I_a$ and $S_t\in I_t$ be the ranges of $S_a$ and $S_t$.
Thus,\footnote{Although demonstrated for specific priors, the proposed method is not limited for these choices and can admit arbitrary priors.}
\begin{equation}
p(S_t,S_a) = \frac{1}{I_a} \times \frac{1}{I_t} .
\label{priorat}
\end{equation}

Let the sensors be placed in $\mathcal{O} \subset \{Z; Z=1,...,N_o \}$ zones, where $N_o$ represents total number of sensors, while the observations are collected at time instances $T_\mathcal{O}=\{t_i\}$.
The observations are used in the Bayesian inference given by Eq. (\ref{BayFin}), with prior defined using Eqs. (\ref{priorNs})-(\ref{priorat}), for rapid source localization and characterization.
In the MCMC implementation of the Bayesian inference, the multizone-CFD simulator is replaced by an appropriate GPE emulator.
Details of the implementation are provided in Algorithm \ref{alg2}.   
To ensure ergodicity, the chain is restarted after initial burn-out period. 
\begin{algorithm}
\caption{MCMC Sampling for GPE based Bayesian Inference}
\label{alg2}
\begin{algorithmic}[1]
\REQUIRE Sensor locations $\mathcal{O}$ and observations $\boldsymbol{Y}_e$ at time instances $T_\mathcal{O}$
\ENSURE $\boldsymbol{\phi}=\{\phi^i\}=\{r_s\in [0,1],r_z\in [0,1],({x_i},{y_i}),S_a,S_t\}$
\STATE Initialize the chain at $\boldsymbol{\phi}=\boldsymbol{\phi}_0$
\FOR{$k=1$ \TO total\_no\_samples}
\FOR{$i=1$ \TO $\text{cardinality}(\boldsymbol{\phi})$}
\STATE Generate a random number $\mathcal{U}\in [-1,1]$
\STATE $\phi^i_* = \phi^i_{k-1} + \mathcal{U}$
\ENDFOR
\STATE $a=\text{int}(\phi^1_* \times N_S + 1)$, $b=\text{int}(\phi^2_* \times N_Z + 1)$ and $\boldsymbol{\theta}=\{\phi^i_*;i=3,...,\text{cardinality}(\boldsymbol{\phi})\}$
\FORALL{$c \in \mathcal{O}$}
\STATE Predict contaminant concentration at time instances $\mathcal{T}_\mathcal{O}$ using emulator $\mathcal{E}_{a,b,c}\left(\boldsymbol{x},\boldsymbol{\theta}\right)$
\ENDFOR
\STATE Calculate posterior probability $p(\boldsymbol{\phi}_*)$ using $\boldsymbol{Y}_e$ and emulator prediction in Eq. (\ref{BayFin})
\STATE Calculate acceptance probability
\begin{equation}
A\left(\boldsymbol{\phi}_*,\boldsymbol{\phi}_{k-1}\right) = \min \left\{1, \frac{p(\boldsymbol{\phi}_*)}{p(\boldsymbol{\phi}_{k-1})} \right\}
\end{equation}
\STATE Generate a uniform random variable $\mathcal{U}$
\IF{$\mathcal{U} \leq A\left(\boldsymbol{\phi}_*,\boldsymbol{\phi}_{k-1}\right)$}
\STATE $\boldsymbol{\phi}_{k} = \boldsymbol{\phi}_*$
\ELSE
\STATE $\boldsymbol{\phi}_{k} = \boldsymbol{\phi}_{k-1}$
\ENDIF
\ENDFOR
\end{algorithmic}
\end{algorithm}
\section{Results and Discussion}
\begin{figure}[h]
\begin{center}
\includegraphics[trim = 30mm 40mm 30mm 30mm, clip, width=3.0in, height=2.85in]{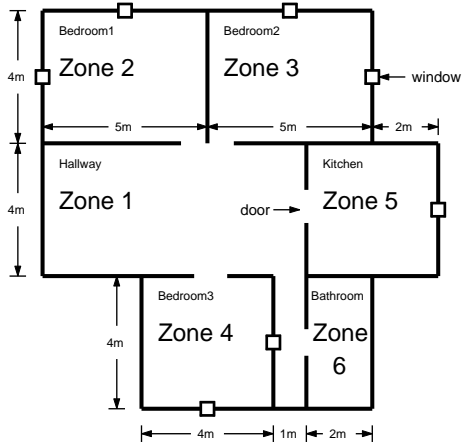}
\end{center}
\vspace*{-.15in}
\caption{Plan of the building}
\label{build_plan}
\end{figure}

Efficacy of the proposed method is demonstrated for localization and characterization of a hypothetical pollutant release in a seven room building.
The building plan is shown in Figure \ref{build_plan}.
Case study is carried out for a single storey 3m high building with one hallway, three bedrooms, a bathroom, a kitchen and a 1m wide open passage.
Rooms are connected internally by doors, while each bedroom is connected to the outside environment by two windows each.
Further, the hallway is connected to the outside environment by a main door.
At the time of contaminant release, all the doors and windows are assumed to be open.
Outside temperature is assumed to be $20^{o}C$ with the wind blowing at 3 m/s.

\subsection{GPE for Multizone-CFD}
\begin{figure}[h!]
  \begin{center}
   \begin{tabular}{l}
   \subfigure [Comparison at points used for emulator] {\includegraphics[width=3.0in, height=1.5in] {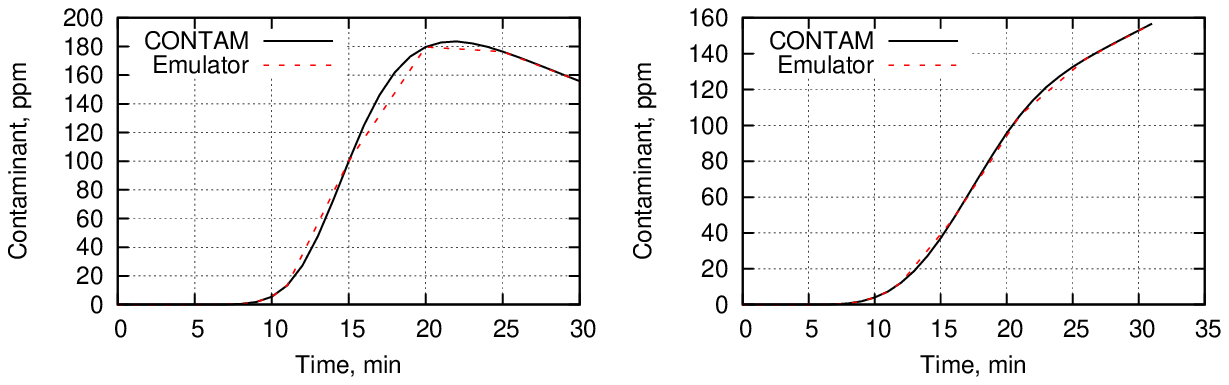}} \\
   \subfigure [Comparison at points not used for emulator] {\includegraphics[width=3.0in, height=1.5in] {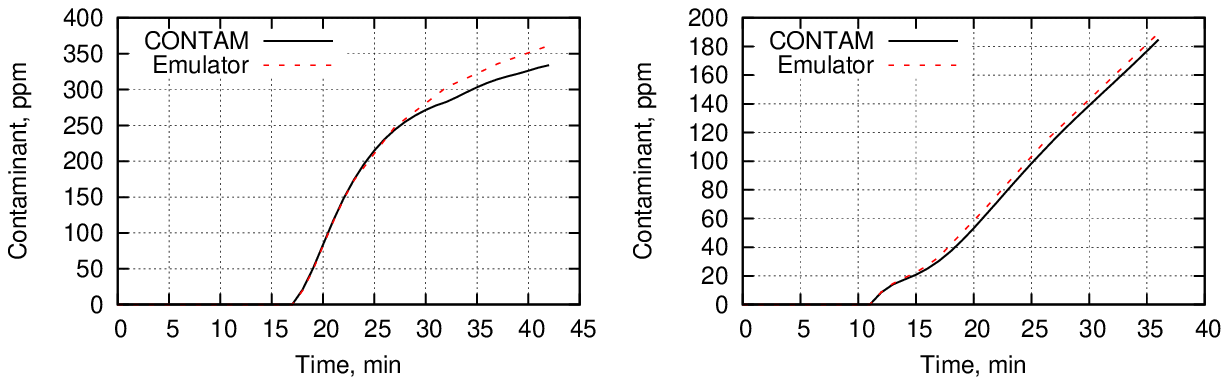}}
   \end{tabular}
  \end{center}
 \caption{Comparison of multizone-CFD and emulator predictions. Figure a) shows comparison at some of the $\boldsymbol{\theta} \in \boldsymbol{S}$. Figure b) shows comparison at $\boldsymbol{\theta} \notin \boldsymbol{S}$.}
\label{emul_comp}
\end{figure}

To build an emulator for the multizone-CFD simulator, an initial set of 121 design points is selected using Latin hypercube sampling \cite{McKay79,Munholland96,HeltonSAND0417}.
For each design point, contaminant concentration at five temporal locations (i.e. $q_1=5$) in the interval of one minutes, starting from one minute after the source activation, is used as a set of initial simulator outputs $\boldsymbol{D}^{ini}_1$.
Conditional on $\boldsymbol{D}^{ini}_1$,  correlation length parameters $\boldsymbol{\lambda}$ are estimated by maximizing Eq. (\ref{postLambda}).
In the present work, Complex Box method \cite{CompBox} is used for optimization.
To avoid local optima, the optimizer is repeatedly run for pre-determined number of times and the best point amongst the resultant optima is chosen as an estimate of $\boldsymbol{\lambda}$.
The initial set of 121 design points is further augmented by sequentially selecting 29 points as described in the Algorithm \ref{alg1}.
The resultant set of 150 design points, $\boldsymbol{S}$, is used to build the GPE.
For each design point from $\boldsymbol{S}$, a second set of simulator outputs, $\boldsymbol{D}_2$, is created by using contaminant concentration values at five temporal locations ($q_2=5$) in the interval of four minutes, starting from $q_1+1$.
Conditional on $\boldsymbol{\lambda}$, $\hat{\mathcal{B}_1}$ and $\hat{\Sigma_1}$ are estimated using $\boldsymbol{D}_1$, while $\hat{\mathcal{B}_2}$ and $\hat{\Sigma_2}$ are estimated using $\boldsymbol{D}_2$.
Fate and transport of the contaminant for first five minutes after the source activation is reconstructed by using estimates of the emulator conditional on $\boldsymbol{D}_1$.
The long term contaminant fate and transport for six minutes onwards from the source activation is reconstructed using estimates of the emulator conditional on $\boldsymbol{D}_2$ along with Eq. (\ref{predTime}).
Figure \ref{emul_comp} shows comparison of transient contaminant concentration obtained using the proposed method with multizone-CFD simulator.

\subsection{Source Localization and Characterization with Full Sensor Network}
Efficacy of the proposed Bayesian framework for rapid source localization and characterization is investigated for a release of contaminant in Hallway (zone 1).
For the present test case, two sources are assumed to be activated at time $T=18$min, with each source releasing carbon monoxide ($CO$) at a rate of 0.09 g/s.
Inside the Hallway, source 1 is located at $(4.0, 1.36)$, while source 2 is located at $(1.44, 3.6)$.
Sensors are assumed to be presented in six zones (zones 1--6, except in Passage, zone 7).
All the sensors are assumed to be collaborating with each other.
Sensor measurement is simulated by running the multizone-CFD with specified source characteristics and location.
Transient multizone-CFD prediction in the time-step of 1 min is used as sensor observations, while experimental uncertainty in each sensor observation is assumed to be $1\%$.
Total 5 data points per sensors (i.e., 5 mins of data) is used for source localization and characterization.
Note that for the present test case, all the zones are connected with the Hallway, thus the contaminant is detected by the sensors in all the zones.
Bayesian inference is used after collecting sensor data for five minutes.
To investigate the efficacy of the proposed method, the Bayesian inference is also implemented using the direct MCMC sampling, where the integrated multizone-CFD model is used in the Metropolis-Hastings algorithm (in Algorithm \ref{alg_metr}) to sample from the posterior distribution.
Total 20,000 samples are collected after burnout period of 10,000 samples.
The resultant posterior distribution is compared with the posterior distribution obtained using the proposed method.
Table \ref{zone_source_prob} summarizes posterior probability of source located inside a given zone and the posterior probability of number of active sources.
For the present test case, the method infers zone and number of sources accurately with probability one.
\begin{table}[h]
\begin{center}
  \caption{Posterior Probabilities of Room \& Number Identification} \vspace*{.05in}
\begin{tabular}{llllllll}
\hline
~ & \multicolumn{7}{l}{Posterior probability of sources in a zone, $p(Z)$} \\  \hline
~ & 1 & 2 & 3 & 4 & 5 & 6 & 7 \\ \hline
Direct MCMC & 1 & 0 & 0 & 0 & 0 & 0 & 0 \\
GPE based MCMC & 1 & 0 & 0 & 0 & 0 & 0 & 0 \\
\hline
~ & ~ & \multicolumn{6}{l}{Posterior probability of no. of sources, $p(S_N)$} \\ \hline
~ & ~ & \multicolumn{2}{l}{1} & \multicolumn{2}{l}{2} & \multicolumn{2}{l}{3} \\ \hline
\multicolumn{2}{l}{Direct MCMC} & \multicolumn{2}{l}{0} & \multicolumn{2}{l}{1} & \multicolumn{2}{l}{0} \\
\multicolumn{2}{l}{GPE based MCMC} & \multicolumn{2}{l}{0} & \multicolumn{2}{l}{1} & \multicolumn{2}{l}{0} \\ \hline
\end{tabular}
\label{zone_source_prob}
\end{center}
\end{table}

\begin{figure}[h!]
\begin{center}
 \begin {tabular}{l l}
    \subfigure [Using direct MCMC]{\includegraphics[width=3.0in, height=2.85in] {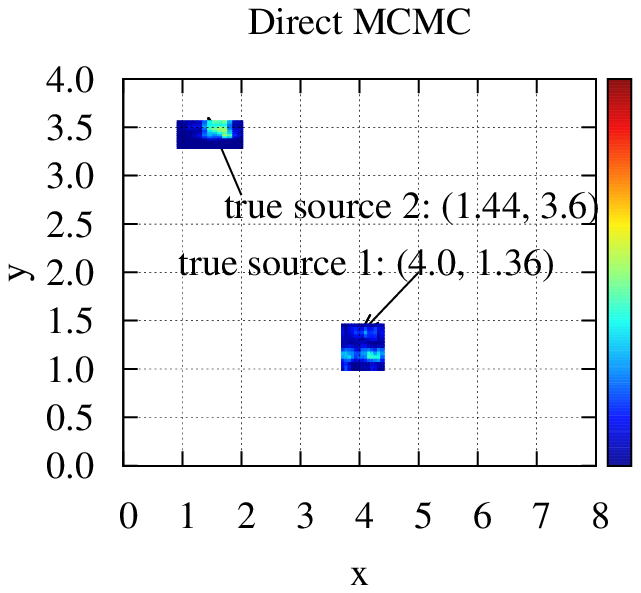}} &
    \subfigure [Using GPE-based MCMC]{\includegraphics[width=3.0in, height=2.85in] {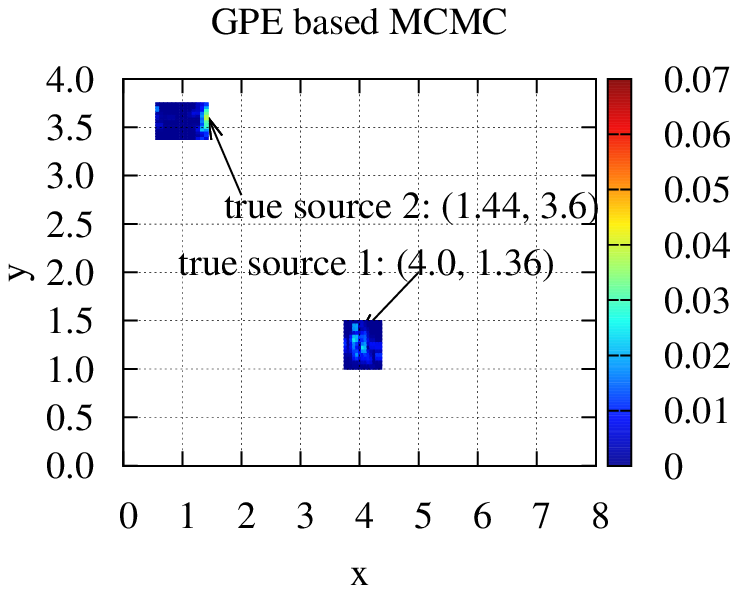}} \\
   \end{tabular}
\end{center}
\vspace*{-.15in}
\caption{Conditional posterior probability distribution of source location}
\label{loc_prob_full}
\end{figure}

The posterior probability contours of source locations obtained using the direct MCMC sampling is shown in Figure \ref{loc_prob_full} (a), while, Figure \ref{loc_prob_full} (b) shows the posterior probability contours obtained using the proposed method.
Actual location of the sources is also indicated in the figure.
From the figure, it may be concluded that the method accurately infers the source location with high probability.
Further, the posterior probability contour obtained using the proposed method matches closely with the direct MCMC sampling.


\begin{figure}[t!]
  \begin{center}
  \begin {tabular}{l l}
    \subfigure []{\includegraphics[width=3.0in, height=2.85in] {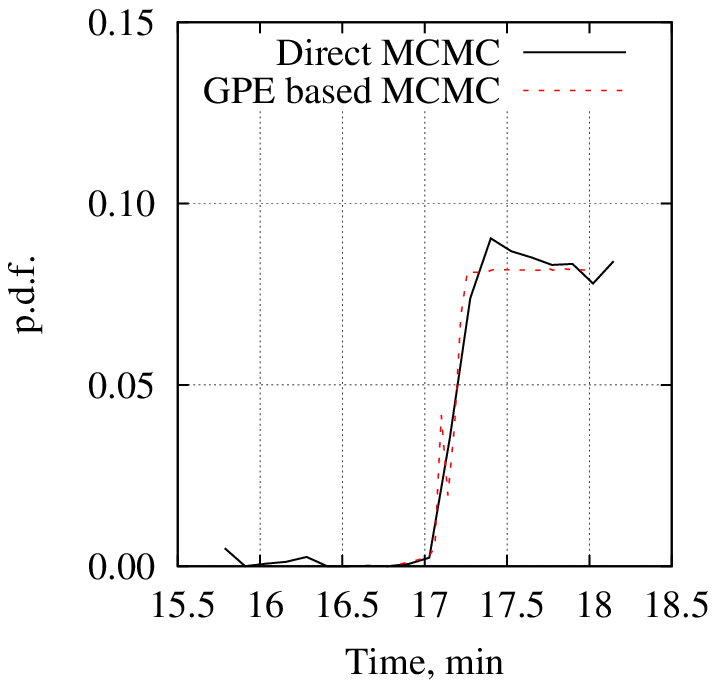}} &
    \subfigure []{\includegraphics[width=3.0in, height=2.85in] {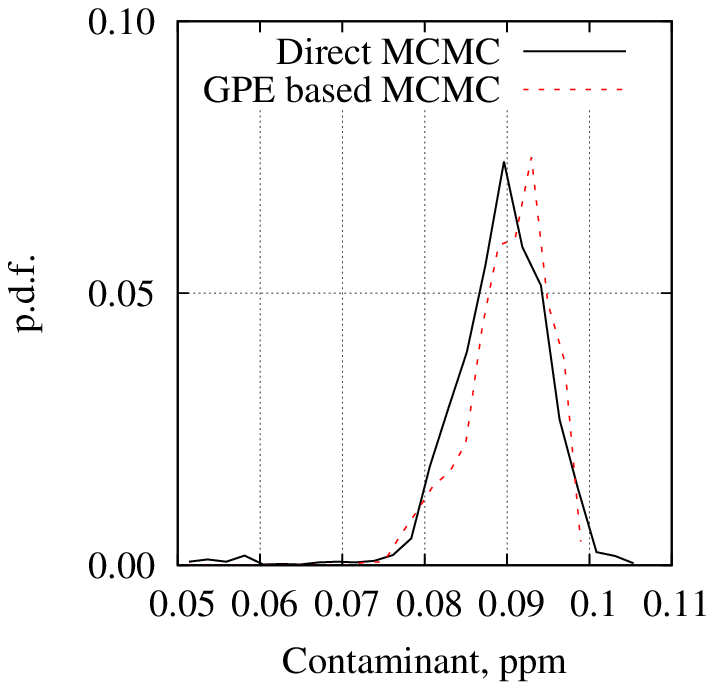}} \\
   \end{tabular}
  \end{center}
 \caption{(a) posterior probability distribution of time of source activation (for the test case, source is activated at 18 min); (b) posterior probability distribution of amount of contaminant released (for the test case, 0.09 g/s CO contaminant is released).}
\label{source_charact_full}
\end{figure}

Figure \ref{source_charact_full} shows posterior probability distribution of the time of source activation and the amount of contaminant release by each source.
Posterior probability distributions obtained using the proposed method and the direct MCMC sampling are shown in the figure, which are found to match closely with each other.
Posterior probability distribution of the release time is a non-symmetric one-sided distribution with high probability near the time of detection and rapidly decreasing away from the detection time, which is similar to the exponential distribution.
Note that this behaviour is expected as the sensors detect contaminants quickly after the source activation, thus the posterior probability near the detection time is high.
Further, as the contaminant accumulates over time, probability of source release at earlier time is low.
Posterior probability of contaminant amount release is symmetric with high probability near $0.09$, which is a true value of contaminant amount release.

\subsection{Effect of Varying Number of Sensors}
\begin{figure}[t!]
  \begin{center}
  \begin {tabular}{l l}
    \subfigure []{\includegraphics[width=3.0in, height=2.85in] {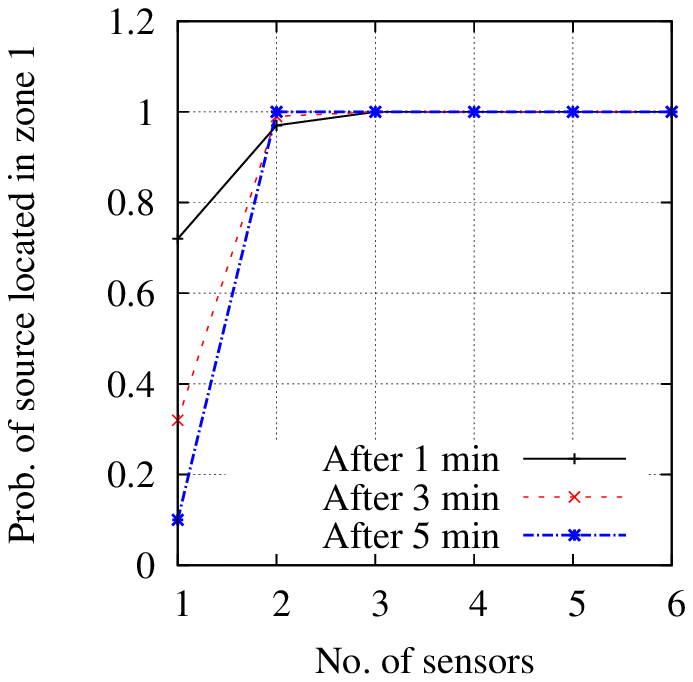}} &
    \subfigure []{\includegraphics[width=3.0in, height=2.85in] {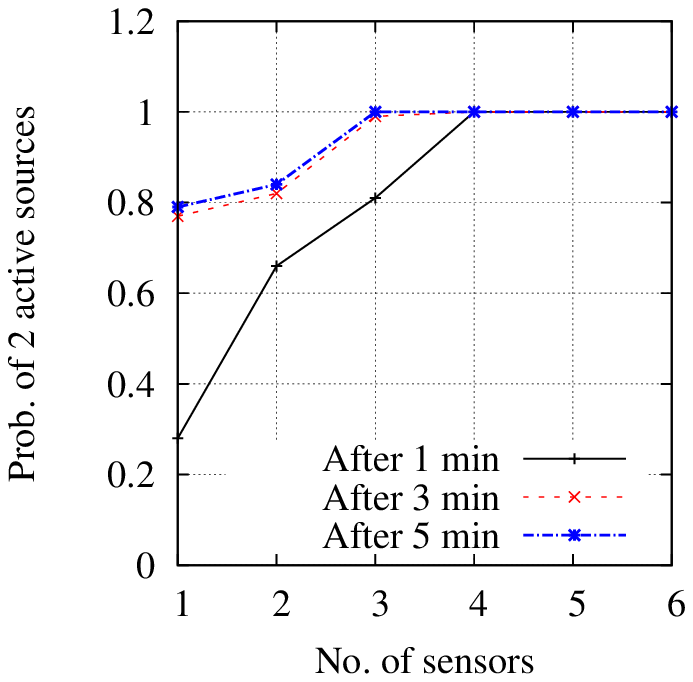}} \\
   \end{tabular}
  \end{center}
 \caption{Inference results with varying number of sensors: (a) posterior probability sources active in zone 1; (b) shows posterior probability of 2 active sources. }
\label{source_charact_varying_ns}
\end{figure}

In this subsection, the proposed method is implemented using the different number of sensors and sensor data points.
All the test cases are presented for two active sources $(S_N=2)$ in the Hallway (zone 1), activated at $S_t = 18~ \text{mins}$ and releasing the carbon monoxide at the rate of $S_a = 0.09$ g/s.
Sensors are assumed to collect the data in the interval of one minute.
The test cases are presented using observations after one minute (1 data point), three minutes (3 data points) and five minutes (5 data points).
All the observations are used concurrently for the Bayesian inference.
Figure \ref{source_charact_varying_ns} (a) shows the posterior probability of sources located in zone 1 using different number of sensors.
When the Bayesian inference is implemented after one minute, the correct zone is inferred with high probability using observation from one sensor, which itself is located in zone 1.
As the number of sensors increases, the posterior probability of sources located in zone 1 increases, with the method inferring the correct zone with probability 1 when three or more sensors are used.
However, when the proposed method is implemented using three or five minutes of data, the correct zone (zone 1) is inferred with low probability, while, the posterior probability is $p(Z=1) = 1$ when the observations from two sensors are used.
As pointed out earlier, zone 1 is connected to other zones, thus, the contaminant released in zone 1 disperses to all the other zones.
Hence, when the contaminant is observed in more than one zone, the posterior probability, $p(Z=1)$, increases rapidly.

Figure \ref{source_charact_varying_ns} (b) shows the posterior probability of $S_N=2$.
For a given number of sensors, the posterior probability, $p(S_N=2)$,increases with increase in the number of observations used.
The inference obtained using 3 minutes and 5 minutes of data matches closely with each other.
When the Bayesian inference is implemented using 1 observation (after 1 minute of data), the number of active sources is inferred correctly with probability one when observations from four or more sensors is used.
However, when 3 or 5 minutes of data is used, number of active sources is inferred with probability one using three or more sensors.

\begin{figure}[t]
\begin{center}
\includegraphics[width=3.0in, height=2.85in]{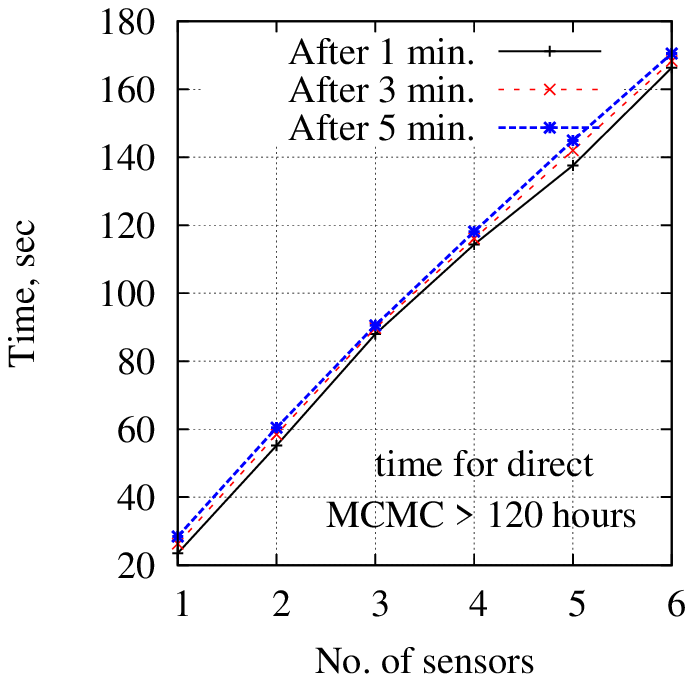}
\end{center}
\vspace*{-.15in}
\caption{Comparison of computational time for different sensor networks and data points}
\label{comp_time_all_varying_ns}
\end{figure}

Figure \ref{comp_time_all_varying_ns} shows computational time for the proposed method.
The computational time for the direct MCMC implementation is also indicated in the figure.
All the test cases are implemented on a desktop computer with Intel Core i5 CPU.
The computational time of the implementation is obtained using a FORTRAN intrinsic routine \emph{cpu\_time}.
The implementation of direct MCMC method, when 5 minutes of data observed by sensors in 6 zones is used, takes more than 120 hours of computational time.
The computational cost of the proposed method is significantly lower than the direct MCMC, demonstrating the possibility of real-time rapid source localization and characterization.
The computational cost of the proposed method increases linearly with the number of sensors used, however, the computational cost does not increase noticeably with the number of observations used.
Note that for each sensor, a separate emulator needs to be evaluated, however, each evaluation of the emulator provide the complete reconstruction of the transient contaminant concentration.
Thus the computational cost increases with the increase in number of sensors used, whereas, increase in the computational cost is minimal for increased number of observations.

\subsection{Inference with Dynamic Incremental Sensor Network}
Results presented in the previous subsection demonstrates the need for a collaborative sensor network for accurate inference.
Computational cost of the proposed method increases with the number of sensors used, limiting the number of sensors for rapid source localization.
However, to ensure that the contaminant is detected in any zone, placement of a sensor in each zone is necessary.
This subsection investigates the proposed method for a possible dynamic sensor network.
The network consists of the sensors placed in six zones (only excluding the passage).
In the event of contaminant detection by any of the sensor, one minute of data of the single sensor is used to infer the posterior probability of zone location and characteristics.
The sensor subsequently requests next three minutes of data from the two adjacent zones with non-zero probability of the source presence, while, the resultant three minutes of observations from three sensors is used for the Bayesian inference.
In the final step, five minutes of observations from all the six sensors are used.
Note that posterior probability of each step is used as a prior for the next step.
The implementation of the resultant sensor network is explained in the Algorithm \ref{sens_net}:
\begin{algorithm}
\caption{Dynamic Collaborative Sensor Network}
\label{sens_net}
\begin{algorithmic}[1]
 \STATE Let the contaminant be detected by the sensor in zone $j$ at time $t$
 \STATE Define $\mathcal{O}=\{j\}$ and $T_\mathcal{O}=\{t+1\}$
 \STATE Specify priors given by Eqs. (\ref{priorNs})-(\ref{priorat})
 \STATE Use Algorithm \ref{alg2} to sample from the posterior distribution \label{post_1}
 \STATE $k=1$
 \FOR{$i=1$ \TO no\_of\_zones}
  \IF{$P(i)\neq 0$}
  \STATE $k=k+1$
  \STATE $\mathcal{O}_{k}=i$
  \ENDIF
\ENDFOR
 \FOR{$i=1$ \TO $3$}
 \STATE ${T_\mathcal{O}}_i=t+1+i$
 \ENDFOR
\STATE Use posterior distribution \ref{post_1} as prior
\STATE Use Algorithm \ref{alg2} to sample from the posterior distribution \label{post_2}
\STATE Define $\mathcal{O}=\{j; j=1,...,6\}$ and $T_\mathcal{O}=\{t+4+i; i=1,...,5 \}$
\STATE Use posterior distribution \ref{post_2} as prior
\STATE Use Algorithm \ref{alg2} to sample from the posterior distribution \label{post_2}
\end{algorithmic}
\end{algorithm}

\begin{figure}[h!]
  \begin{center}
   \begin{tabular}{l l l}
   \subfigure []{\includegraphics[width=2.0in, height=1.5in] {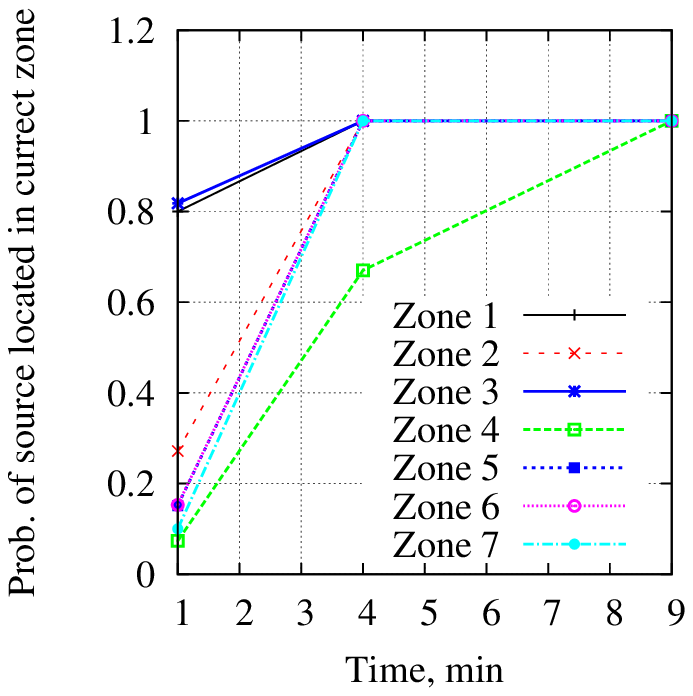}} &
   \subfigure []{\includegraphics[width=2.0in, height=1.5in] {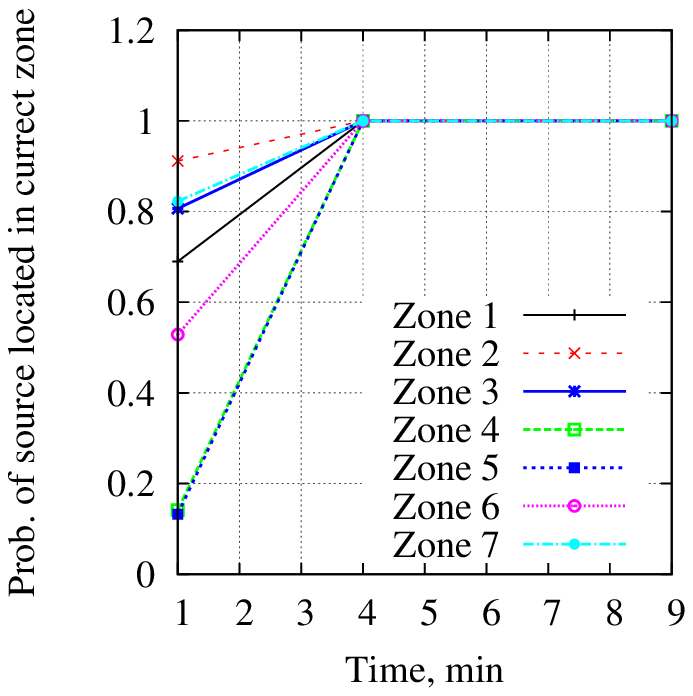}} &
   \subfigure []{\includegraphics[width=2.0in, height=1.5in] {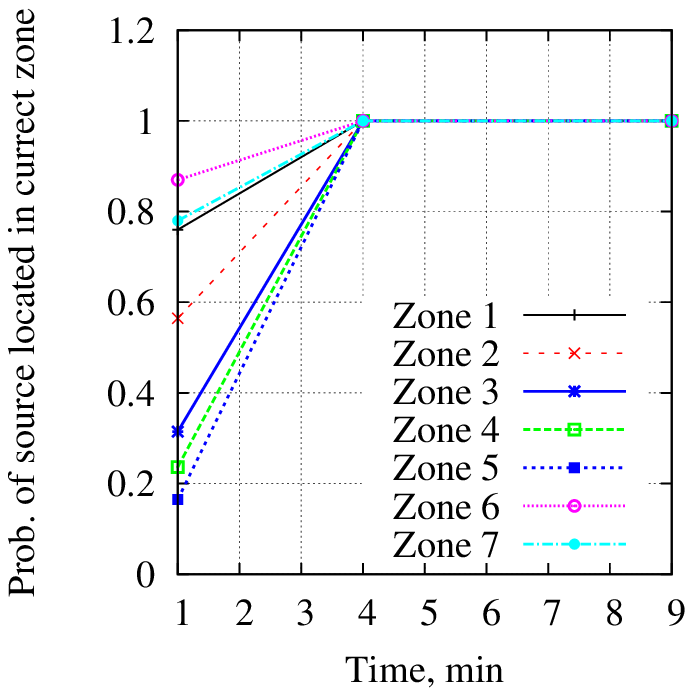}} \\

   \subfigure []{\includegraphics[width=2.0in, height=1.5in] {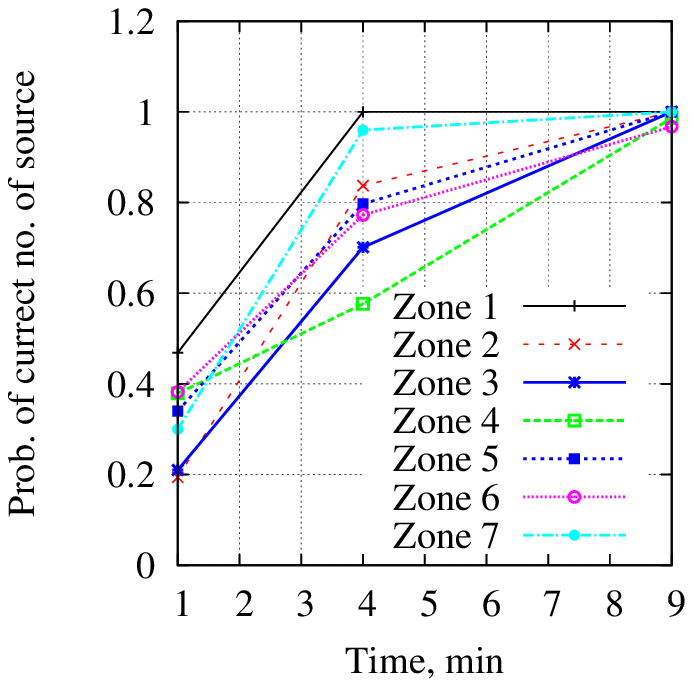}} &
   \subfigure []{\includegraphics[width=2.0in, height=1.5in] {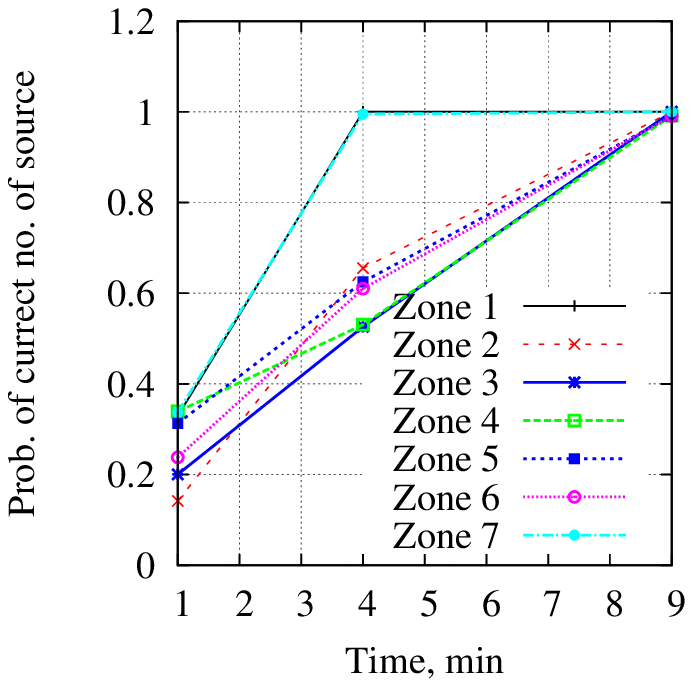}} &
   \subfigure []{\includegraphics[width=2.0in, height=1.5in] {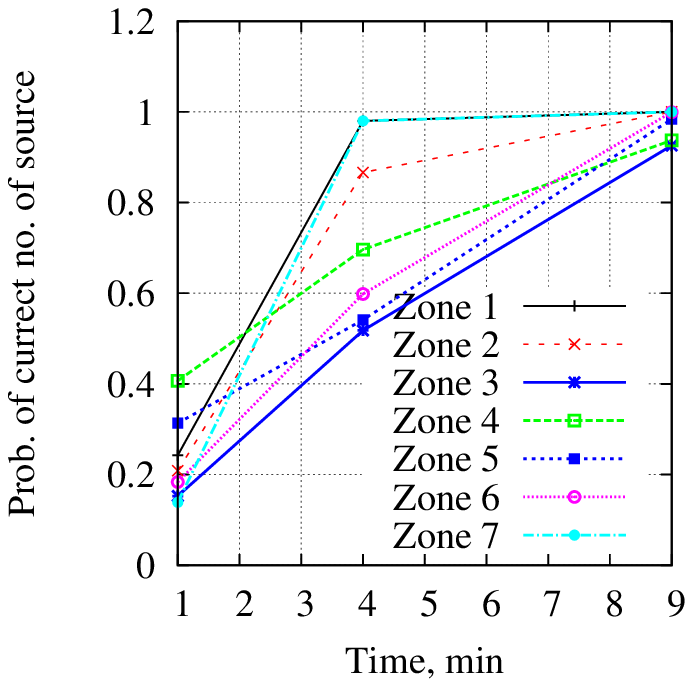}}
   \end{tabular}
  \end{center}
 \caption{Source localization using collaborative sensor network. Results are shown for sources located in different zones. Top row shows posterior probability of true zone and bottom row shows posterior probability of true number of sources. Left column shows results for one active source, middle column shows results for two active sources and right column shows results for three active sources.}
\label{source_local}
\end{figure}

In this subsection, efficacy of the collaborative sensor network, described in Algorithm \ref{sens_net}, is investigated for different test cases.
Figure \ref{source_local} compares the effect of number of sources active in different zones on the source localization.
The results are presented for sources activated at 18 minutes releasing 0.09 g/s of carbon monoxide.
The top row of Fig. \ref{source_local} shows posterior probability of the `true' zone, $p(Z)$, where each test case represent sources located in different zones.
Left row show results for one active source, middle row show results for two active sources, while the right row show the results for three active sources.
The bottom row show similar results for posterior probability of the `true' number of sources, $p(S_N)$.
As can be observed from the figure, the `true' zone of active sources is correctly located by the collaborative sensor network after four minutes (i.e., after using observations from three sensors), for all the test cases except when the single source is active in zone 4, which is inferred correctly with probability one after nine minutes.
The `true' number of active sources is inferred after four minutes with varying posterior probability, however after nine minutes, the `true' number of active sources is inferred with $p(S_N)$ approaching one for all the test cases.
Thus, the collaborative sensor network explained in Algorithm \ref{sens_net} can accurately localize the source using the proposed method with four minutes of sensor observations from three collaborating sensors, while the number of active sources is also inferred accurately after using the nine minutes of data from six sensors.

\begin{figure}[h!]
  \begin{center}
   \begin{tabular}{l l l}
   \subfigure [True time of release = 6 min]{\includegraphics[width=2.0in, height=1.5in] {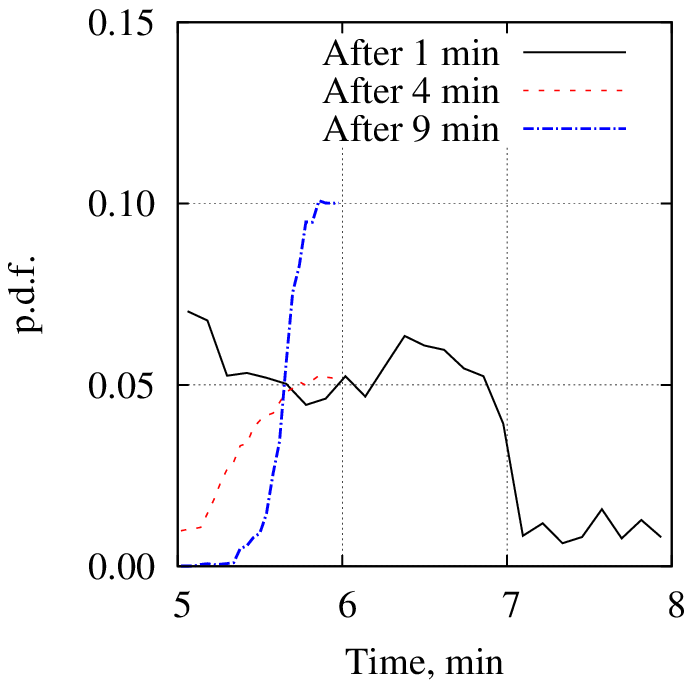}} &
   \subfigure [True time of release = 12 min]{\includegraphics[width=2.0in, height=1.5in] {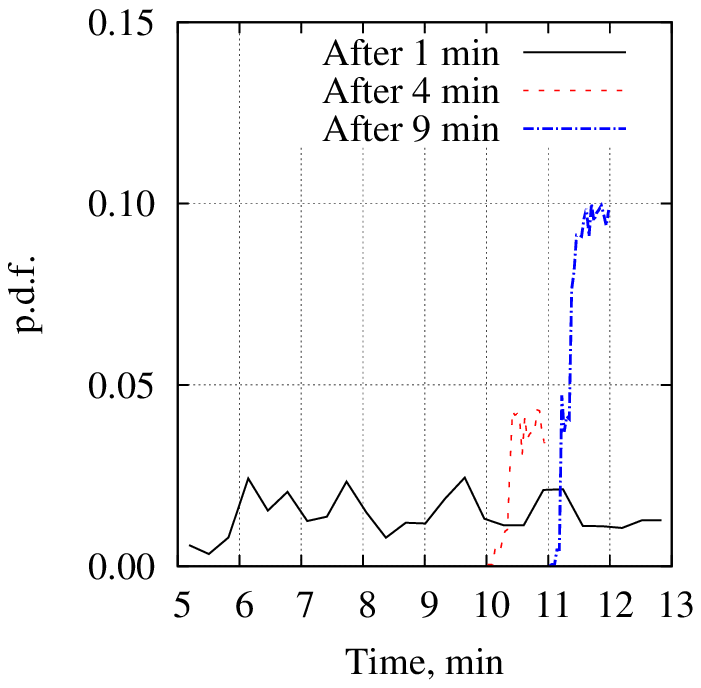}} &
   \subfigure [True time of release = 18 min]{\includegraphics[width=2.0in, height=1.5in] {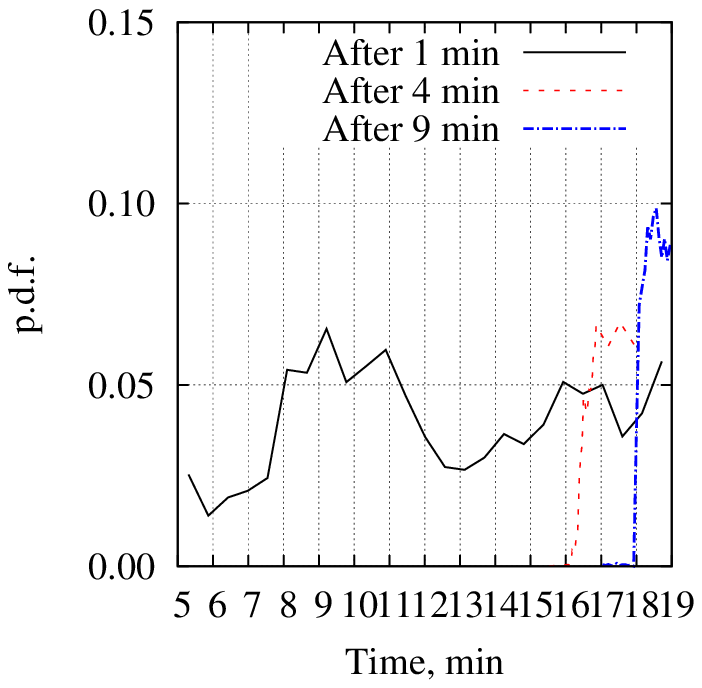}} \\

   \subfigure [True amount of source = 0.06 g/s]{\includegraphics[width=2.0in, height=1.5in] {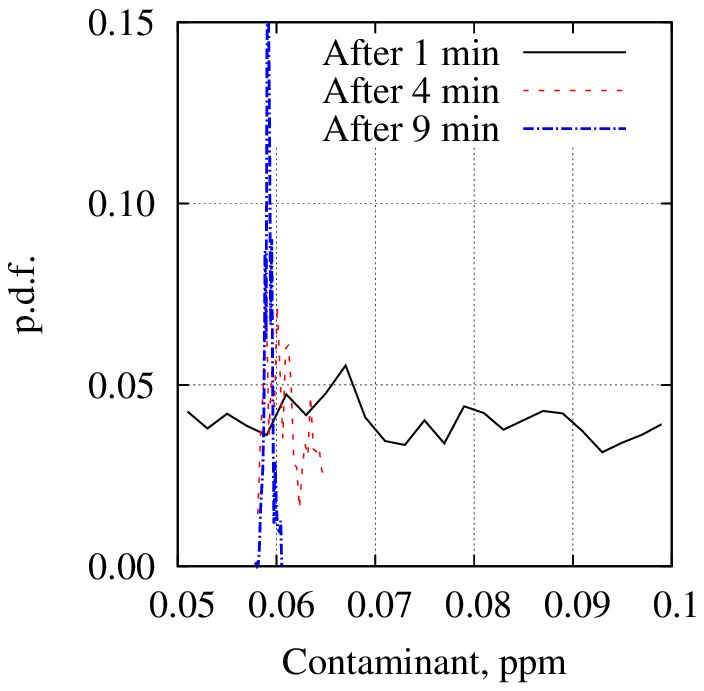}} &
   \subfigure [True amount of source = 0.08 g/s]{\includegraphics[width=2.0in, height=1.5in] {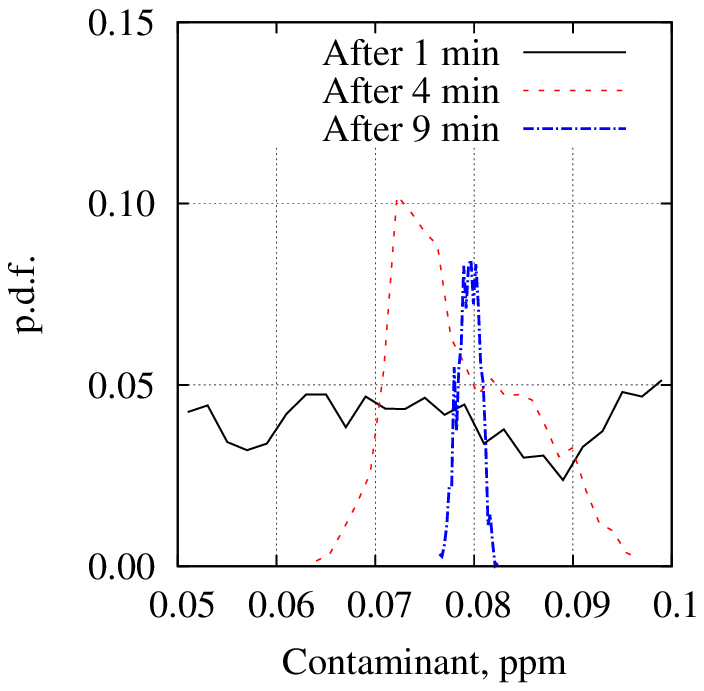}} &
   \subfigure [True amount of source = 0.09 g/s]{\includegraphics[width=2.0in, height=1.5in] {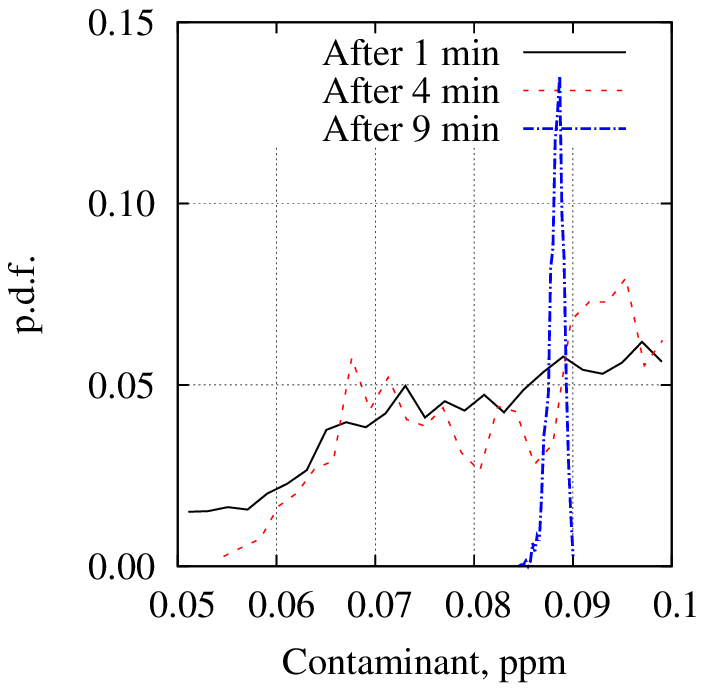}}
   \end{tabular}
  \end{center}
 \caption{Source characterization collaborative sensor network. Top row shows posterior probability of time of source activation and bottom row shows posterior probability of amount of source released.}
\label{source_charact}
\end{figure}

Figure \ref{source_charact} investigates the efficacy of the proposed method, using the collaborative sensor network (Algorithm \ref{sens_net}), to infer the time and amount of contaminant release.
The results are presented for $Z=1$ and $S_N=2$.
Top row of the Figure \ref{source_charact} shows the posterior probability distribution of time of source activation $S_t$, with $S_a=0.09 \text{g/s}$, while, the bottom row shows the posterior probability of $S_a$ when $S_t=18 \text{mins}$.
As the more observations are used from increasing number of sensors, the posterior probability distribution becomes narrow around the `true' value.
For all the test cases presented, time and amount of contaminant release is inferred correctly with high probability after four minutes using three collaborating sensors, while the probability of the `true' values increases when the network of six collaborating sensors is used after nine minutes.
The results presented in this subsection have demonstrated the feasibility of using the proposed method for rapid source localization and characterization for a possible dynamic incremental sensor network.
The method can similarly be applied for investigating other sensor networks.

\section{Concluding Remarks}
This paper have presented a Gaussina proecess emulator (GPE)-based Bayesian framework for rapid contaminant source localization and characterization in the indoor environment.
The framework can be used with a computationally expensive integrated multizone-CFD model.
The framework approximates the multizone-CFD model using a GPE during the pre-event detection stage, which is used for Bayesian inference of the source location and characteristics after the contaminant detection by the sensors.
The framework provide methodology for rapid localization and characterization of multiple sources.
In conjunction with the rapidly advancing digital and sensor technologies, the framework can be used for planning the evacuation and the source extinguishing strategies in an indoor building environment in view of sudden contaminant release.
The framework can also be used to test different sensor networks and investigate the performance tradeoffs.

In the present paper, efficacy of the framework have been investigated for an hypothetical contaminant release in a single storey seven room building.
The posterior distributions of the uncertain parameters obtained using the proposed method are found to match closely with the direct MCMC implementation, at a significantly lower computational cost. Performance and the robustness of the proposed method have been investigated for a dynamic incremental sensor network. 
Various test cases presented in the paper have demonstrated the robustness of the proposed method, although in a limited sense for one of a possible sensor network. 
In future, authors propose to investigate the presented approach as an inference machine for informative sensor planning. 
 

\bibliographystyle{elsarticle-num}



\end{document}